%% file: main.tex
\documentclass[a4paper,11pt, superscriptaddress]{quantumarticle}
\pdfoutput=1
\usepackage[utf8]{inputenc}
\usepackage[english]{babel}
\usepackage[T1]{fontenc}
\usepackage[numbers,sort&compress]{natbib}
\usepackage{svg,times}
\usepackage{color}
\usepackage{graphicx}
\usepackage{dcolumn}
\usepackage{bm}
\usepackage{physics}
\usepackage{hyperref}
\usepackage{url}
\usepackage{amsmath}
\usepackage{tikz-cd}
\usepackage{notes2bib}
\usepackage{ dsfont }
\usepackage{svg}
\usepackage{verbatim}
\usepackage{amssymb}
\usepackage{amsfonts}              
\usepackage{amsthm}                
\usepackage{mathtools}
\usepackage{qcircuit}
\usepackage[normalem]{ulem}
\usepackage{xcolor}

\usepackage{tikzit}
\input{sample.tikzstyles}

\DeclarePairedDelimiterX{\barpair}[2]{(}{)}{%
  #1\;\delimsize\|\;#2%
}
\theoremstyle{definition}
\newtheorem{theorem}{Theorem}
\newtheorem{prop}[theorem]{Proposition}

\newtheorem{conjecture}[theorem]{Conjecture}
\newtheorem{definition}[theorem]{Definition}
\newtheorem{corollary}[theorem]{Corollary}

\newcommand{\STr}{\mf{Tr}}
\newcommand{\Ad}{\text{Ad}}
\newcommand{\id}{\text{id}}
\newcommand{\mf}[1]{\mathfrak{#1}}

\newcommand{\mcal}[1]{\mathcal{#1}}

\newcommand{\mds}[1]{\mathds{#1}}

\bibnotesetup{note-name    = ,use-sort-key = false} 
\begin{document}

\title{Quantum operations with the time axis in a superposed direction}

\author{Seok Hyung Lie}
\affiliation{%
 School of Physical and Mathematical Sciences, Nanyang Technological University, 21 Nanyang Link, 637371, Singapore
}%
\author{M.S. Kim}
\affiliation{%
QOLS, Blackett Laboratory, Imperial College London, London, SW7 2AZ, UK.
}%
\date{23-05-2023}

\begin{abstract}
\onecolumn
In the quantum theory, it has been shown that one can see if a process has the time reversal symmetry by applying the matrix transposition and examining if it remains physical. However, recent discoveries regarding the indefinite causal order of quantum processes suggest that there may be other, more general symmetry transformations of time besides the complete reversal. In this work, we introduce an expanded concept of matrix transposition, the generalized transposition, that takes into account general bipartite unitary transformations of a quantum operation's future and past Hilbert spaces, allowing for making the time axis definitely lie in a superposed direction, which generalizes the previously studied `indefinite direction of time', i.e., superposition of the forward and the backward time evolution. This framework may have applications in approaches that treat time and space equally like quantum gravity, where the spatio-temporal structure is explained to emerge from quantum mechanics. We apply this generalized transposition to investigate a continuous generalization of perfect tensors, a dynamic version of tracing out a subsystem, and the compatibility of multiple time axes in bipartite quantum interactions. Notably, we demonstrate that when a bipartite interaction is consistent with more distinct local temporal axes, there is a reduced allowance for information exchange between the two parties in order to prevent causality violations.
\end{abstract}

\maketitle

\section{Introduction}
The arrow of time has been one of the central topics of physics. Although the unique direction of time propagation from the past to the future is so natural for us living in the macroscopic and classical world, the fundamental laws of nature governing the microscopic world appear to be symmetric with respect to the reversal of time direction. There are some attempts to explain the emergence of the arrow time with thermodynamics argument in the classical realm.

Recently, Chiribella and Liu studied the time reversal symmetry of quantum processes \cite{chiribella2020quantum}. In Ref. \cite{chiribella2020quantum}, it is shown that the most appropriate mathematical representation of the input-output reversion is the matrix transposition, and the quantum processes that are consistent with both directions of time propagation correspond to unital quantum channels. However, the input-output reversion may not be the most general symmetry transformation of the temporal structure of quantum processes considering recent developments in indefinite causal structures of quantum processes \cite{chiribella2013quantum}.

Especially, in some approaches to quantum theory of spatio-temporal structure of the universe like the quantum gravity theory, the spacetime is also treated on the same footing with other quantum objects. In such theories, the existence of the unique flow of time is not assumed. Some approaches explain that the time emerges only after some subspace of the whole Hilbert space of the universe is identified as a `clock' to provide quantized time parameter \cite{page1983evolution}. In this picture, there is no immediate reason to expect that there is a unique well-defined direction of time obeyed by every quantum system in the universe, as there is an ambiguity in the choice of a clock system, known as the clock ambiguity \cite{albrecht2008clock,albrecht2012clock,marletto2017evolution}. In other words, when interpreted as quantum systems, the distinction between future and past systems is not so clear, and the partition between them need not be unique.

These observations suggest the possibility of altering the direction of temporal direction, not just within a given axis -forward and backward or their superpositions as considered in Ref. \cite{chiribella2013quantum, chiribella2020quantum}- but also through the transformation of the direction of temporal axis \textit{itself}. In this work, we develop a generalization of the approach of Chiribella and Liu \cite{chiribella2020quantum} by introducing the \textit{generalized transposition}, which generalizes the conventional matrix transposition and study its applications and implications in various contexts such as tensor network picture of quantum events, perfect tensors and information exchange within bipartite quantum interactions.

\subsection{Notation} \label{subsec:Not}
Without loss of generality, we sometimes identify the Hilbert space $H_X$ corresponding to a quantum system $X$ with the system itself and use the same symbol $X$ to denote both. We will denote the dimension of $X$ by $|X|$. For any system $X$, $X^{\otimes n}$ represents the tensor product of $n$ copies of $X$, and when we need to refer to one copy of it, we denote it by $X',X''$ etc. In other words, $X'$ is a copy of $X$ with the same dimension, i.e., $|X|=|X'|$. When there are many systems, all the systems other than $X$ are denoted by $\bar{X}$. However, the trivial Hilbert space will be identified with the field of complex numbers and will be denoted by $\mds{C}.$ The identity operator on system $X$ is denoted by $\mds{1}_X$ and the maximally mixed state is denoted by $\pi_X=|X|^{-1}\mds{1}_X$. The space of all bounded operators acting on system $X$ is denoted by $\mf{B}(X)$, the real space of all Hermitian matrices on system $X$ by $\mf{H}(X)$. The set of all unitary operators in $\mf{B}(X)$ is denoted by $\mf{U}(X)$. For any $M\in \mf{B}(X)$, we let $\Ad_M$ be $\Ad_M(K):=MKM^\dag$. For any matrix $M$, $M^T$ is its transpose with respect to some fixed basis, and for any $M\in\mf{B}(X\otimes Y)$, the partial transpose on system $X$ is denoted by $M^{T_X}$. The Schatten $p$-norm of an operator $X$ is defined as $\norm{X}_p:=\Tr[(X^\dag X)^{p/2}]^{1/p}=\{\sum_i (s_i(X))^p\}^{1/p}$ where $s_i(X)$ is the $i$-th largest singular value of $X$. The (Uhlmann) fidelity between two quantum states is defined as $F(\rho,\sigma):=\norm{\sqrt{\rho}\sqrt{\sigma}}_1^2$.

The space of all linear maps from $\mf{B}(X)$ to $\mf{B}(Y)$ is denoted by $\mf{L}(X, Y)=\mf{B}(\mf{B}(X),\mf{B}(Y))$ and we will use the shorthand notation $\mf{L}(X):=\mf{L}(X, X)$. The set of all quantum states on system $X$ by $\mf{S}(X)$ and the set of all quantum channels (completely positive and trace-preserving linear maps) from system $X$ to $Y$ by $\mf{C}(X , Y)$ with $\mf{C}(X):=\mf{C}(X , X)$. Similarly we denote the set of all quantum subchannels (completely positive trace non-increasing linear maps) by $\tilde{\mf{C}}(X,Y)$ and $\tilde{\mf{C}}(X):=\tilde{{\mf{C}}}(X,X)$. We denote the identity map on system $X$ by $\id_X$. For any completely positive map $\mcal{N}=\sum_i\Ad_{K_i}$, we define its transpose as $\mcal{N}^T:=\sum_i\Ad_{K_i^T}$.

$J_{XX'}^\mcal{N}$ is the Choi matrix of $\mcal{N} \in \mf{L}(X)$ defined as $J_{XX'}^\mcal{N}:=\mcal{N}_X(\phi^+_{XX'})$ where $\phi^+_{XX'}=\dyad
{\phi^+}_{XX'}$ is a maximally entangled state with $\ket{\phi^+}_{XX'}=|X|^{-1/2}\sum_i\ket{ii}_{XX'}$. The mapping $J:\mf{L}(X)\to\mf{B}(X\otimes X')$ defined as $J(\mcal{M}):=J_{XX'}^\mcal{M}$ itself is called the Choi-Jamio{\l}kowski isomorphism \cite{choi1975completely,jamiolkowski1972linear}. Unnormalized state $\sum_i\ket{ii}_{XX'}$ will be denoted by $\ket{\Gamma}_{XX'}$. We call a linear map from $\mf{L}(X)$ to $\mf{L}(Y)$ a \textit{supermap} from $X$ to $Y$ and denote the space of supermaps from $X$ to $Y$ by $\mf{SL}(X,Y)$ and let $\mf{SL}(X):=\mf{SL}(X,X)$. Supermaps preserving quantum channels even when it only acts on a part of multipartite quantum channels are called \textit{superchannel} \cite{chiribella2008transforming,gour2019comparison,chiribella2009theoretical,burniston2020necessary,bisio2019theoretical,chiribella2013quantum,gour2020dynamical} and the set of all superchannels from $X$ to $Y$ is denoted by $\mf{SC}(X,Y)$ and we let $\mf{SC}(X):=\mf{SC}(X,X)$.   We say a superchannel $\Omega\in\mf{SC}(X)$ is \textit{superunitary} if there are $U_0$ and $U_1$ in $\mf{U}(X)$ such that $\Omega(\mcal{N})=\Ad_{U_1}\circ\mcal{N}\circ\Ad_{U_0}$ for all $\mcal{N}\in \mf{L}(X)$.

We define the `Choi map' $\mds{J}[\Theta]\in\mf{L}(X\otimes X',Y\otimes Y')$ of supermap $\Theta\in\mf{SL}(X,Y)$ in such a way that the following diagram is commutative:
\begin{equation}
    \begin{tikzcd}[baseline=\the\dimexpr\fontdimen22\textfont2\relax]
    \mf{L}(X)\ar[d,"J"]\ar[r,"\Theta"] & \mf{L}(Y) \ar[d,"J"] \\
    \mf{B}(X\otimes X')\ar[r,"{\mds{J}[\Theta]}"]& \mf{B}(Y\otimes Y')
  \end{tikzcd}.
\end{equation}
Similarly, we define the inverse of the Choi map $\mds{J}^{-1}[\mcal{N}]\in\mf{SL}(X,Y)$ of a linear map $\mcal{N}\in\mf{L}(X\otimes X',Y\otimes Y')$ in such a way that the following diagram commutes:
\begin{equation}
    \begin{tikzcd}[baseline=\the\dimexpr\fontdimen22\textfont2\relax]
    \mf{L}(X)\ar[d,"J"]\ar[r,"{\mds{J}^{-1}[\mcal{N}]}"] & \mf{L}(Y) \ar[d,"J"] \\
    \mf{B}(X\otimes X')\ar[r,"{\mcal{N}}"]& \mf{B}(Y\otimes Y')
  \end{tikzcd}.
\end{equation}

\section{Generalized transpose}
Imagine that an experimenter observes a quantum system evolving with the passage of time. The process may appear to have well-defined input and output systems for the experimenter. However, how can one be sure that the quantum system experiences the same passage of time with the classical experimenter outside of the system?  This seemingly obvious question is actually highly nontrivial considering the fact that time is not a universal parameter shared by all the systems but a quantity that should be observed with a physical mean as one can see from the difficulty in constructing a satisfactory quantum clock \cite{peres1980measurement, buvzek1999optimal}. The possibility of superposition of multiple time evolutions has been studied since decades ago \cite{aharonov1990superpositions}. Especially, with the recent development of indefinite causal structure of quantum systems \cite{ball2017world}, it is evident that there are no a priori reasons to assume that a quantum process has a unique temporal axis.

\begin{figure}[t]
    \centering
    \includegraphics[width=.5\textwidth]{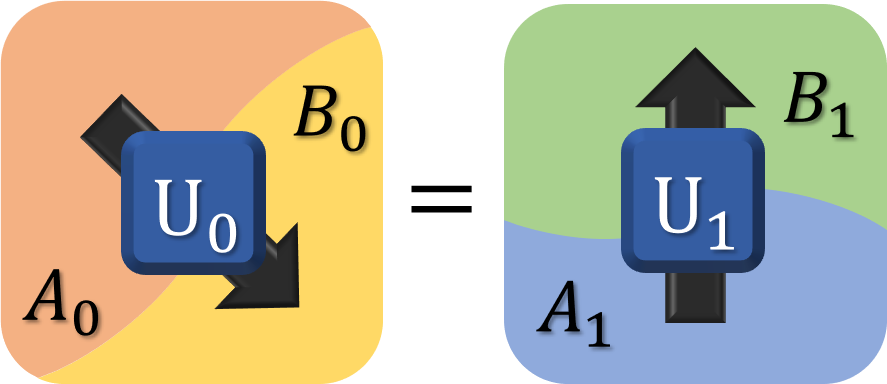}
    \caption{Consider a unitary operation $U_0:A_0\to B_0$. One can interpret $A_0$ and $B_0$ as the past and the future Hilbert spaces of a quantum system and $U_0$ as its time evolution. However, If $A_0\otimes B_0 = A_1\otimes B_1$ with $W$ being the bipartite unitary operator connecting two different tensor decompositions, then $U_0$ can be interpreted as $U_1$ from $A_1$ to $B_1$ if $U_1=U_0^{T[W]}$.}\label{fig:equivalence}
\end{figure}

Nevertheless, if an experimenter can prescribe a valid description of a given quantum process, e.g., a completely positive trace preserving (CPTP) map, or, a quantum channel, then we can conclude that at least one temporal structure, that is, the one the experimenter follows, is compatible with the given quantum process. However, by no means that should be the unique temporal structure compatible with the process. A quantum process connects input and output systems, but the distinction between them is made from the perspective of the experimenter; There could be other partitionings of the input-output joint system into an alternative input-output system pair (See FIG. \ref{fig:equivalence}.) One could consider it a new type of symmetry a quantum process could have. Then, a natural question follows: How can one describe the corresponding symmetry transformation? (See Sec. \ref{sec:spt} for extended discussion on the necessity of studying temporally indefinite quantum states.)

In this work, we construct such a symmetry transformation by generalizing the matrix transposition. The input-output inversion of quantum operation is given as the transpose operation $M\mapsto M^T$ \cite{chiribella2020quantum}, which can be understood as the rotation by 180$^\circ$ in tensor diagram, i.e.
\begin{equation} \label{eqn:trans}
    \input{trans.tikz},
\end{equation}
whose action on a quantum channel with Kraus operators $\qty{K_n}$ is given as
\begin{equation}
    \sum_n K_n \rho K_n^\dag \mapsto \sum_n K_n^* \rho K_n^T.
\end{equation}
(Adjoint operation $\dag$, unlike transposition, does not function properly when applied to a local system of multipartite quantum channel, essentially because complex coefficient of a bipartite state does not belong to a single system. See Ref.  \cite{chiribella2020quantum}.) Geometrically, one can interpret the transpose operation as flipping the direction of time. However, since there are ways to rotate a diagram other than the rotation by $180^\circ$, one could naturally become curious after seeing (\ref{eqn:trans}) if there is a way to express the general transformation of the direction of the \textit{time axis itself}, not just flipping the time direction for a given time axis as in Ref. \cite{chiribella2020quantum}. In this section, we propose such a generalization.

The unitary operator is the complex generalization of the orthogonal matrix, hence it generalizes the action of rotating to complex Hilbert spaces. First, we observe that we can stretch and curve the wires in (\ref{eqn:trans}) to transform it into
\begin{equation} \label{eqn:TWtrans}
    \input{TWtrans.tikz}.
\end{equation}
Here, the crossing wires in the right hand side can be interpreted as the swapping operator $F:=\sum_{ij}\dyad{i}{j}\otimes\dyad{j}{i}$, which is a unitary operator.

On the other hand, we also have the following expression,
\begin{equation} \label{eqn:TWident}
    \input{TWident.tikz}.
\end{equation}
Here, we can say that the swapping operator in (\ref{eqn:TWtrans}) is replaced by the identity operator. Comparing (\ref{eqn:TWtrans}) and (\ref{eqn:TWident}), one could understand transpose operation as the exchange of future and past Hilbert space. We can naturally guess that if we substitute them with a general bipartite unitary operator, we can get a generalization of transpose operation. Therefore, we define \textit{the generalized transposition} $T[W]$ for each bipartite unitary operator $W\in\mf{U}(A^{\otimes2})$ which maps any $M \in \mf{B}(A)$ to $M^{T[W]}$ defined in the following way,
\begin{equation}
  \input{TW0.tikz},
\end{equation}
where the arrow next to each box indicates the flow of time, or the direction from input to output. Concretely, $M^{T[W]}$ is defined as, with a fixed basis $\{\ket{i}\}$ of $A$, 
\begin{equation} \label{eqn:PTe}
    M^{T[W]}:=\sum_{i,j} (\mds{1}\otimes\bra{j})W(M\ket{i}\otimes\dyad{i}{j}).    
\end{equation}

It is remarkable that generalized transpositions are the only kind of linear maps that preserve the Hilbert-Schmidt inner product, i.e., every linear map $\Phi$ on a matrix algebra satisfying
\begin{equation} \label{eqn:HSinner}
    \Tr[\Phi(A)^\dag\Phi(B)]=\Tr[A^\dag B],
\end{equation}
for arbitrary operators $A,B$ is a generalized transposition. It follows that that generalized transpositions preserve the Schatten 2-norm, i.e. $\|X^{T[W]}\|_2=\|X\|_2$ for every $X$ and $W$. We remark that superunitary operations (See Sec. \ref{subsec:Not}) are also a special case of the generalized transpose operation. It is because whenever $W$ is a product unitary operator, say, $W=W_1\otimes W_2^T$ with $W_1,W_2\in\mf{U}(A)$, $M^{T[W]}=W_1 M W_2$. Note that although we defined $T[W]$ only for unitary operators $W$, obviously it can be extended to arbitrary operators and $T$ is linear for general inputs.

This generalization immediately enables the `fractional transposition'. It means that, for any angle $\theta$, one can now express the rotation of tensor diagram of a given operator by $\theta$. For any $\theta\in\mds{R}$, the operator given as $F(\theta):=F^{\theta/\pi}=e^{-i\theta/2}(\cos(\theta/2)\mds{1}_{AA'}+i\sin(\theta/2) F_{AA'})$ is also unitary. Hence, the fractional transposition $T(\theta):=T[F(\theta)]$ can be defined, and it outputs the superposition of an operator and its time-inverted version, i.e.,
\begin{equation}
    M^{T(\theta)}=e^{-i\theta/2}(\cos\frac{\theta}{2} M + i\sin\frac{\theta}{2} M^T).
\end{equation}
For instance, for Pauli operators $X$ and $Z$ on $\mds{C}^2$, we have $(XZ)^{T({\pi/2})}=(e^{-i\pi/4}XZ+e^{i\pi/4}ZX)/\sqrt{2}$. As it is expected from rotation transformations, fractional transpositions have the group property : $M^{T({\theta+\phi})}=(M^{T(\theta)})^{T(\phi)}=(M^{T(\phi)})^{T(\theta)}$. Interestingly, if $M$ is symmetric, i.e. $M^T=M$, then it is also invariant under fractional transpositions, i.e. $M^{T(\theta)}=M$ for all $\theta$.

Once we defined how generalized transposition acts on matrices, we can define the corresponding action of generalized transposition on quantum channels, or more generally, on linear maps. We can define the supermap $\mf{T}[W]$ given as
\begin{equation} \label{eqn:PTsch}
    \mf{T}[W](\mcal{N})(\sigma):= |A|^2 \Tr_{BA'}[ (\mds{1}_A\otimes\phi_{BA'}^+)(\Ad_W(J_{AB}^\mcal{N})\otimes\sigma_{A'})],
\end{equation}
for any $\sigma\in\mf{B}(A')$ and $\mcal{N}\in\mf{L}(A)$. For the sake of brevity, we will sometimes use the notation $\mcal{N}^{T[W]}:=\mf{T}[W](\mcal{N})$. This seemingly complicated definition of superchannel $\mf{T}[W]$ is given in this way so that $(\Ad_M)^{T[W]} =\Ad_{M^{T[W]}}$. From this, one can easily see that if
\begin{equation}
    \mcal{N}(\rho)=\sum_n c_n K_n\rho K_n^\dag,
\end{equation}
with complex numbers $c_n\in\mds{C}$, then
\begin{equation}
    \mcal{N}^{T[W]}(\rho)=\sum_n c_n K_n^{T[W]}\rho K_n^{T[W]\dag}.
\end{equation}

One important distinction should be made at this point. Although they share the same mathematical form, the generalized transposition defined here is for quantum processes, not quantum states. Transposition acting on density matrices is important for testing NPT entanglement \cite{peres1980measurement, horodecki1996necessary}, but does not necessarily have the operational meaning as the reversal of input-output systems of a quantum process.

Given our tool for describing symmetry transformations of temporal structures in quantum processes, we can now define the compatibility of a quantum process with multiple temporal structures using generalized transposition. It is a direct generalization of \textit{bidirectional operations} corresponding to the conventional transposition considered in Ref. \cite{chiribella2020quantum}.

\begin{definition}
    A quantum channel $\mcal{N}$ is compatible with a generalized transposition $T[W]$ when $\mcal{N}^{T[W]}$ is also a channel.
\end{definition}

As closed quantum systems evolve with time via unitary operations, it is considered that unitary operations are basic building blocks of time evolution of quantum systems. We immediately get the following result on the generalized transposition of unitary operations by simply observing that $\Tr\circ\Ad_{U^{T[W]}}=\Tr$ is equivalent to $U^{T[W]\dag}U^{T[W]}=\mds{1}$.

\begin{prop} \label{prop:utou}
    If a unitary operation $\mcal{U}$ is compatible with $T[W]$, then $\mcal{U}^{T[W]}$ is also a unitary operation.
\end{prop}

Formally it is obviously possible to generalize the generalized transpose even further by letting the unitary operation $W$ to be a general quantum channel, but we focus on unitary cases in this work. It is mainly because allowing for irreversible quantum operations seems to go against the interpretation of the generalized transpose as a coordinate transformation of future and past Hilbert spaces, not an active joint evolution of future and past systems, albeit a probabilistic implementation through quantum teleportation is possible as it is for the transposition \cite{chiribella2020quantum}.

One subclass of generalized transpositions of special interest is that of \textit{unital generalized transpositions}. A bipartite unitary operator $W$ has the maximally entangled state $\sum_i \ket{i}\ket{i}$ as an eigenvector with eigenvalue 1 if and only if $T[W]$ is unital, since for such $W$, $\mds{1}^{T[W]}=\mds{1}$. Note that a generalized transposition $T[W]$ is unital if and only if it is trace-preserving, i.e.  $\Tr M^{T[W]}=\Tr M$ for all $M$. For example, every fractional transposition, including the usual transposition, is trace-preserving and unital. Since unital generalized transpositions preserve the identity operator, they have the operational meaning as a transformation that preserves `no event', which is desirable for a transformation of time axis. Imagine that there exists a film that recorded no event happening at all; it is natural to expect that playing it forward, backward, or even in quantumly weird direction of time makes no difference.

One possible problem of the definition of generalized transposition is that it may be too general to represent the transformation of tensor product decomposition because there are multiple bipartite unitary operators $W\in\mf{U}(AB)$ that preserve the tensor product structure of the Hilbert space.
We can observe that the nonlocal properties of $W$ in the definition of generalized transposition $T[W]$ correspond to the properties of $T[W]$ as a transformation of temporal structure, as $W$ can be interpreted as a bipartite interaction between ``future'' and ``past'' systems.

Therefore we consider the equivalence class of bipartite unitary operators that are similar through local unitary operators, i.e., $\langle W \rangle=\{(u_1\otimes u_2)W(v_1\otimes v_2): u_1,v_1\in \mf{U}(A), u_2,v_2\in\mf{U}(B) \}$, so that every unitary operator in the same class has the same nonlocal properties. Note that every operator in the same equivalence class transforms the tensor product structure in the same way. This leaves the problem of choosing a good representative from each equivalence class, and from its desirable properties, we hope to choose a bipartite unitary operator that induces a unital generalized transposition. When is it possible?

We say that a bipartite unitary operator \textit{preserves maximal entanglement (ME)} when it maps at least one maximally entangled state to a maximally entangled state. This definition when combined with the definition of the equivalence class of locally similar bipartite unitary operators yields the following result.

\begin{prop}
    There is a unital generalized transposition $T[V]$ with $V\in\langle W\rangle$ if and only if $W$ preserves ME.
\end{prop}

Notably, every two-qubit unitary operator preserves ME \cite{ye2004entanglement, kraus2001optimal} as there are always at least four maximally entangled states that form an orthonormal basis remaining maximally entangled after the action of the unitary operator. Hence, it is conjectured that every bipartite unitary operator preserves ME, even in higher dimensions \cite{puchala2015certainty,brahmachari2022dual}. This conjecture can be compactly stated with the generalized transposition.

\begin{conjecture}[UBB, \cite{puchala2015certainty,brahmachari2022dual}] \label{conj:UBB}
    For every $W\in \mf{U}(AA'),$ there exists at least one pair $(U,V)$ of unitary operators in $\mf{U}(A)$  such that
    \begin{equation}
        U^{T[W]}=V.
    \end{equation}
\end{conjecture}

Especially, there is a numerical evidence of this conjecture that there is an iterative algorithm that finds a sequence of pairs of quantum states that converge to a pair of maximally entangled states related by a given bipartite unitary operator \cite{brahmachari2022dual}. If Conjecture \ref{conj:UBB} is true, then we can always pick a representative that yields the unital generalized transposition from each equivalence class of locally similar bipartite unitary operators. It is equivalent to that the only nontrivial effect of generalized transposition to a transformation comes from its unital part, and all the other effects can be understood as unitary operation applied before and after the transformation in question.

This conjecture, when limited to the class of controlled unitary operators, is equivalent to the following problem.

\begin{conjecture}[UBB-CU (Controlled unitary)]
    For every set of $d$ unitary operators $\qty{U_i}$ on $d$-dimensional Hilbert space $\mcal{A}$, there is an orthonormal basis $\qty{\ket{\psi_i}}$ of $\mcal{A}$ such that $\qty{U_i\ket{\psi_i}}$ is also an orthonormal basis of $\mcal{A}$.
\end{conjecture}

One can see that this conjecture is equivalent to the UBB conjecture for controlled unitary operators of the form $\sum_i \dyad{i}\otimes U_i$ from the fact that arbitrary maximally entangled pure state must have an expression of the form of $d^{-1/2}\sum_i \ket{i}\otimes \ket{\psi_i}$ for some orthonormal basis $\qty{\ket{\psi_i}}$. Namely, after the action of the unitary operator, the state is transformed into $d^{-1/2}\sum_i \ket{i}\otimes U_i\ket{\psi_i}$, thus $\qty{U_i \ket{\psi_i}}$ must be an orthonormal basis, too.

When expressed in this form, it is evident that the UBB-CU conjecture is also equivalent to its classical counterpart. In other words, when it is promised that a random index $i$ will be picked and accordingly the unitary operator $U_i$ will be applied to the quantum system $A$ which contains the memory of the index value $i$, it is natural to conjecture that there exists a deterministic process, represented by a unitary process $\ket{i} \mapsto \ket{\psi_i}$, that prepares a quantum state $\ket{\psi_i}$ that retains the memory of the index $i$ after the action of $U_i$. The UBB-CU conjecture supposes that exactly such a process always exists for any set of $\qty{U_i}$.

One simple example is the case of the generalized transposition corresponding to the CNOT gate, i.e.,  $W=\dyad{0}\otimes\mds{1}+\dyad{1}\otimes X$. The Hadamard gate $H=\dyad{+}{0}+\dyad{-}{1}$ is compatible with $T[W]$, as one can see from
\begin{equation}
   H^{T[W]}=XH, 
\end{equation}
where $X$ is the Pauli-X operators. One can unitalize $T[W]$ by substituting $W$ with $W':=W(\mds{1} \otimes X^{1/2} H)$, so that ${\mds{1}}^{T[W']}=\mds{1}$.

We remark that even if not every bipartite unitary operator preserves ME, in light of Proposition $\ref{prop:utou}$, we could argue that only generalized transpositions corresponding to those preserve ME are relevant for the temporal structure of quantum processes. It is because if a $W$ does not preserve ME, then no two unitary operators are related to each other via the corresponding generalized transposition $T[W]$. However, if one includes the non-unitary quantum channels into the picture, then it is no a priori clear if there are no pairs of quantum channels related by a non-unital generalized transposition. We leave this problem as an open problem.

Note that the generalized transposition is basis-dependent as the conventional transposition is a basis-dependent operation. There are two layers of basis dependency for input and output systems, i.e, the choice of basis $\qty{\ket{i}}$ and $\qty{\ket{j}}$ in (\ref{eqn:PTe}). One could interpret Choosing the unital representation locally similar to a given generalized transposition is eliminating one such basis dependency by equalizing the input and output bases.

Just as the transposition can be applied to a part of multipartite operators to define the partial transpose operation, the generalized transposition can also be applied to a part of multipartite operator. If $M\in\mf{B}(AB)$, then, for arbitrary $W\in\mf{U}(B^{\otimes2})$, the partial generalized transposition $T_B[W]$ is defined as $T_B[W]:=\id_A\otimes T[W]$
\begin{equation}
  \input{TWpart.tikz}.
\end{equation}

Using the partial generalized transposition, we can examine the compatibility of bipartite unitary operators with multiple directions of time of a local system. Assume again that two systems $A$ and $B$ interact through a bipartite unitary operator $V\in\mf{U}(AB)$. This assumption alone has a couple of implications. It assumes that there are two subsystems $A$ and $B$ that can be localized and identified, which stays so even after the interaction. Also, it also implies that $A$ and $B$ appear to share the same time axis during their interaction. However, this need not be the unique description of the direction of time for each system. For example, $B$ might also appear to evolve in the direction given by a generalized transposition $T_B[W]$ in time from  the perspective of $A$. In this case, for interaction $V$ to be consistent with the new flow of time as well, its generalized transpose $V^{T_B[W]}$ also should be unitary. The same argument can be applied to general quantum channels, so we give the following definition of compatibility.

\begin{definition}
    A quantum channel $\mcal{N}\in\mf{C}(AB)$ is compatible with a generalized transposition $T_B[W]$ on $B$ when $\mcal{N}^{T_B[W]}:=(\mf{id}_A\otimes\mf{T}_B[W])(\mcal{N})$ is also a channel.
\end{definition}

For the case of conventional matrix transposition $T$, bipartite unitary operators compatible with $T$ on a subsystem is called to be t-dual or catalytic. (See Sec. \ref{subsec:compstp} for more information.)

Finally, we examine the relation between the compatibility of a quantum process with multiple directions of time and that of its causally neighbouring processes. When seen from a broader perspective, no interactions happen isolated as they are embedded in the network of events (e.g., see FIG. \ref{fig:timedir}). For example, at the very least, experimenter prepares an input state and measures the output state of a given quantum channel.

\begin{figure}[t]
    \centering
    \includegraphics[width=.9\textwidth]{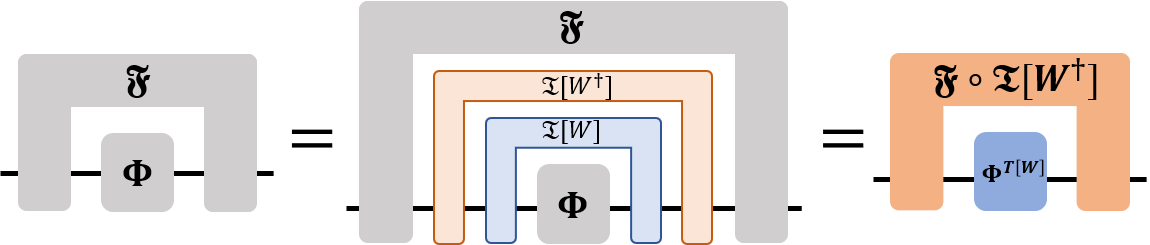}
    \caption{Compatibility of superchannel with the generalized transposition of its input channel.}\label{fig:SupChGenTra}
\end{figure}

We can model the ambient quantum processes as a quantum superchannel since they map a quantum channel to another quantum channel. Therefore, when we examine the consistency of causalities, it is natural to also require the physicality of the causality of the ambient superchannel. If a quantum channel $\Phi\in\mf{C}(A)$ embedded in a superchannel $\mf{F}$ is compatible with a generalized transposition $\mf{T}[W]$, then, for this generalized transposition of $\Phi$ to be consistent with $\mf{F}$ as well, we require that $\mf{F}\circ\mf{T}[W^\dag]$ is also a superchannel, because (See FIG. \ref{fig:SupChGenTra}.)
\begin{equation}
    \mf{F}\circ\mf{T}[W^\dag]\left(\mf{T}[W](\Phi)\right)=\mf{F}(\Phi),
\end{equation}
and because every superchannel with one input register is guaranteed to be physically implementable with a pre-process and a post-process \cite{chiribella2008transforming}. In other words, if one tries to re-interpret a given event in a different decomposition of the spacetime, then the events surrounding it must be consistent as well in that decomposition.

This observation severely restricts which state can be fed into a multipartite unitary operator with multiple compatible temporal axes, as the following Proposition shows.

\begin{prop} \label{prop:comm}
    A state preparation superchannel given as $\mf{P}^\sigma(\mcal{N}):=\mcal{N}(\sigma)$ is compatible with a generalized transposition $T[W]$ of its input channel, i.e., $\mf{P}^\sigma\circ\mf{T}[W^\dag]$ is a superchannel, if and only if there exists a quantum state $\tau$ such that
    \begin{equation} \label{eqn:prepcom}
      W(\mds{1}_{A}\otimes\sigma_{A'}^T)=(\mds{1}_{A}\otimes\tau_{A'}^T)W.
    \end{equation}
\end{prop}
 Note that (\ref{eqn:prepcom}) implies that $\tau$ and $\sigma$ are unitarily similar. Proof can be found in Appendix. From Proposition \ref{prop:comm}, it follows that the preparation of the maximally mixed state is always compatible with an arbitrary generalized transposition of its input channel. (See Sec.\ref{subsec:STr} for a related discussion on factorizable maps.) Especially, for the case of time inversion, corresponding to the transposition, $W$ is the swapping gate and Proposition \ref{prop:comm} implies that $\sigma=\mds{1}/\Tr[\mds{1}]$, hence we have the following Corollary.

\begin{corollary} \label{coro:maxmix}
    The only state preparation compatible with the transposition is the preparation of the maximally mixed state.
\end{corollary}

This result matches with our intuition that no knowledge can propagate backward in time, as feeding a non-maximally mixed states into a quantum process compatible with inverse evolution can lead to retrocausality. See Sec. \ref{subsec:compstp} for the discussion on the constraint Proposition \ref{prop:comm} imposes on the information exchange between two quantum systems through a bipartite quantum channel when there are multiple compatible local temporal directions.

\section{Discussion} \label{sec:consist}
\subsection{Events in spacetime as a tensor network} \label{sec:spt}
\begin{figure}[t]
    \centering
    \includegraphics[width=.5\textwidth]{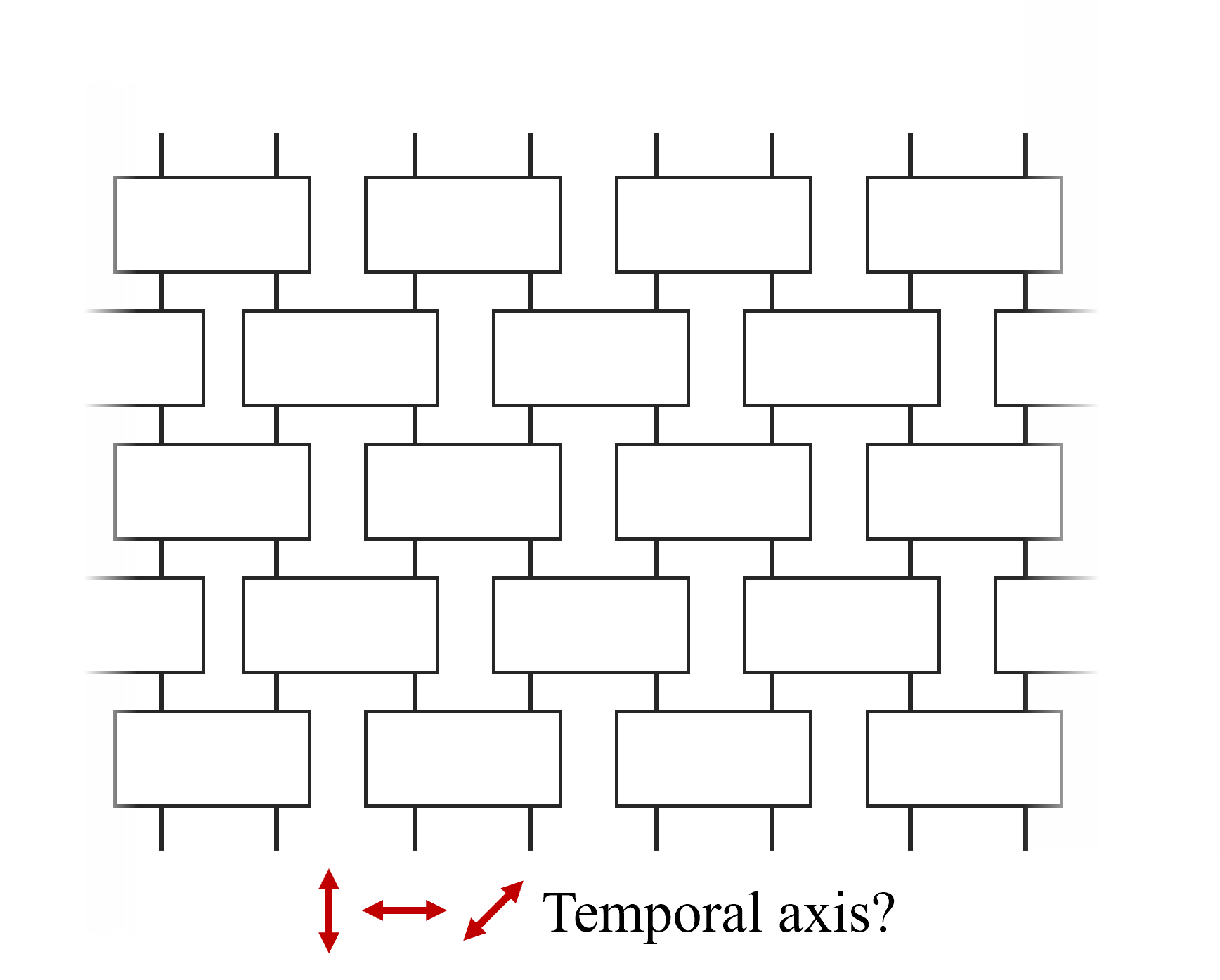}
    \caption{Suppose that the dynamics of a set of quantum systems is given as a tensor network. Without presupposing a spacetime structure, it may not be possible to assign the unique temporal axis of the evolution in the tensor network.}\label{fig:timedir}
\end{figure}
In this Section, we delve into a more detailed discussion on spacetime regions with indefinite causal orders and how the generalized transposition can be used in the context. In conventional quantum mechanics, the quantum state of a quantum system $\ket{\psi(t)}$ at time $t$ can be written as

\begin{equation}
    \ket{\psi(t)}=\left(\prod_{t> t'\geq 0} U_{t'}\right) \ket{\psi(0)},
\end{equation}
where each $U_{t'}$ is the unitary operator describing the time evolution from time $t'$ to time $t'+1$. Moreover, each $U_{t'}$ also can be decomposed into interaction between many subsystems located at $x$, e.g. $U_{t'}=\bigotimes_x U_{(x,t')}$. The dynamics $\ket{\psi(t)}$ went through can be depicted as a tensor network resembling FIG \ref{fig:timedir} where each box is $U_{(x,t')}$. Therefore, once the set of unitary operators $\{U_{(x,t)}\}$ and their connectivity are given, the dynamics of a set of quantum systems is completely decided. In other words, we can consider the dynamics of quantum systems \textit{a net of events} composed of unitary operator which can be interpreted as a tensor. This approach shares the same spirit with the approach known as the \textit{event-universe} \cite{shor2022towards,shor2022dendrographic} understanding the universe as a tree of events except that `events' are unitary evolution, or quantum channels, in this work.

However, what if we do not assume that there is a spacetime with the familiar spatio-temporal structure? What if the existence of the universal axis of time is not given as an additional data outside of the Hilbert space of the universe? There are approaches to quantum gravity in which they treat time, which is often treated as a parameter, on the same footing with space. One of the purposes of theses approaches is to recover time as an emergent entity from quantum theory without supposing its familiar properties as the temporal parameter. Notable examples include those of Cotler \cite{cotler2020toward,cotler2016entangled}, Castellani \cite{castellani2021entropy,castellani2021history} and Dias \cite{dias2021quantum}.

We can consider the following time-neutral model of the Hilbert space of the spacetime. Suppose that there exists a `pre-spacetime' structure $\mcal{S}$, which parametrizes different regions of the to-be spacetime that is not necessarily having familiar spatio-temporal properties. We suppose that the Hilbert space $\mcal{H}$ of a part of or the whole universe can be decomposed into smaller Hilbert spaces $\mcal{H}_s$ corresponding to each point $s$ in  $\mcal{S}$ 
\begin{equation} \label{eqn:history}
    \mcal{H}_{\mcal{S}}=\bigotimes_{s\in\mcal{S}} \mcal{H}_{s}.
\end{equation}
One possible model of pre-spacetime $\mcal{S}$ is the set of points $(\textbf{x},t)$ in the familiar spacetime before the spatio-temporal structure is assigned. In some literature, the Hilbert space of a quantum system including its behavior through the passage of time is called the \textit{history Hilbert space} \cite{isham2003new,cotler2016entangled,cotler2018superdensity}, hence we will use the same nomenclature for $\mcal{H}_{\mcal{S}}$ in (\ref{eqn:history}). However, since we are yet to allocate the temporal parametric role to parameter $t$ in $(\textbf{x},t)$, we stick to the temporally neutral notation $s$ for each region in $\mcal{S}$. Also, for the sake of simplicity, we suppose that the structure of $\mcal{S}$ is discrete by interpreting each $s$ as a region in spacetime rather than a point. Working with continuous tensor product requires the full kit of Fahri-Gutmann path integral formalism \cite{farhi1992functional}, which goes beyond the scope of the current work. Generalization to the continuous regime is an interesting future work.

Assuming the existence of a net of events is not conceptually more demanding than other approaches \cite{carroll2019mad, carroll2021quantum} that assume pre-existing Hamiltonian attached to each Hilbert space. It is immediate from the fact that assuming that a Hamiltonian $H$ governs a quantum system is mathematically equivalent to assuming that the dynamics of the quantum system is described by the unitary operator $U:=\exp{-iHt/\hbar}$. Nevertheless, in this work, we deem the picture of unitary operators and the corresponding quantum channels constituting the history of the universe is conceptually more clear than the picture of Hamiltonians living outside of Hilbert space yet inducing dynamics of quantum systems indirectly.


Additionally, we will work on a plausible assumption that every $\mcal{H}_s$ is finite-dimensional and isomorphic with each other. Assuming isomorphic structure amounts to assuming a sort of translational symmetry of the pre-spacetime $\mcal{S}$ which is usually done in cosmology. Also, there are good reasons to assume that the Hilbert space of the universe is locally finite-dimensional based on the arguments such as the finiteness of the entropy of black holes \cite{cao2017space}.

Once the history Hilbert space is defined, we assume that an event of the universe is given as a tensor $X\in\bigotimes_{s\in\mcal{S}'}\mcal{H}_s$ with some $\mcal{S}'\subseteq\mcal{S}$ connecting different regions of the pre-spacetime $\mcal{S}$. However, we are more familiar with the interpretation of an event as an operator transforms a state into another, hence, whenever we can divide the support Hilbert space of $X$ (we will sometime call this the history Hilbert space of $X$) into two disjoint same-dimensional regions $\mcal{S}_0$ and $\mcal{S}_1$, we tentatively interpret $X$ as an operator between these regions as follows.
\begin{equation}
    X: \bigotimes_{s\in\mcal{S}_0}\mcal{H}_s \to \bigotimes_{s\in\mcal{S}_1}\mcal{H}_s.
\end{equation}
This kind of identifying tensor with operator is often done in the field of quantum scrambler \cite{hosur2016chaos}. However, it may not be possible to interpret $X$ as a time-evolution from $\bigotimes_{s\in\mcal{S}_0}\mcal{H}_s$ to $\bigotimes_{s\in\mcal{S}_1}\mcal{H}_s$, as the input and output systems can be arbitrarily chosen since $X$ was given as a tensor in $\bigotimes_{s\in\mcal{S}_0\bigcup\mcal{S}_1} \mcal{H}_s$ and the operator interpretation was rather arbitrary. At this stage, we only have a set of Hilbert spaces corresponding to each distinguishable points in the `pre-spacetime' whose physical meaning is still unclear and a set of rather abstract `events' given as tensors on the Hilbert spaces.

Now, we hope to study the spatio-temporal structure that emerges from the given network of events. This approach is particularly motivated by the recent results including the HaPPY code \cite{pastawski2015holographic}, where the correspondence between the tensor network and spacetime emerged is explored. Especially, it was shown that the correspondence with tensor network can be extended from that of space-like time slices to the whole spacetime, recently \cite{may2017tensor}.

For the case of pre-spacetime $\mcal{S}$ without a presupposed temporal structure, there need not be a unique axis of time across the whole tensor network. As an example, one can imagine the following interaction $X$ between four regions $s_i$ in $\mcal{S}$.
\begin{equation} \label{eqn:tennet_1}
    \input{tennet_1.tikz}.
\end{equation}
One could consider $X$ as a tensor in $\bigotimes_{i=0}^3\mcal{H}_{s_i}$. However, without a presupposed temporal parameter, there is no \textit{a priori} reason to suppose that there is a causal order between the regions, as it is evident from that the Hilbert space $\bigotimes_{i=0}^3\mcal{H}_{s_i}$ has infinitely many ways to be decomposed into the tensor product of input and output spaces. 

In other words, by changing the basis, the tensor $X$ as a vector in $\bigotimes_{i=0}^3\mcal{H}_{s_i}$ can appear as $UX$ with some unitary operator $U$ on $\bigotimes_{i=0}^3\mcal{H}_{s_i}$. This unitary operation can nontrivially affect the operator interpretation of $X$ (say, from $s_0s_1$ to $s_2s_3$), but the description of the effect can be complicated when one is working with the operator interpretation. In the next Section, we will develop a language that can concisely describe this type transformation of operators.

To recover the temporal structure, we adopt the approach taken in the studies of quantum causal models \cite{allen2017quantum} which assumes that quantum dynamics is fundamentally unitary even when the causal relationship between events is unclear.
Say, if $X$ is a unitary operator from $s_0s_1$ to $s_2s_3$ in (\ref{eqn:tennet_1}), then one could give an interpretation to $X$ as the time evolution from the temporally preceding regions $s_0s_1$ to the temporally succeeding regions $s_2s_3$.  Hence, although there may not exist a unique global time axis in $\mcal{S}$, there still could be a ``local time axis'', given in terms of the decomposition of its supporting Hilbert space into ``future'' and ``past'' subsystems, allowing us to interpret each local interaction as a unitary evolution along some direction in $\mcal{S}$.

Of course, the inverse of a unitary operator is still unitary, there is no canonical distinction between ``future'' and ``past'' yet, which should be made by other means, e.g. the second law of thermodynamics. Therefore, the terms like axis or direction of time should be understood with this symmetry in consideration. It is somehow similar to that both $\ket{\psi}$ and $-\ket{\psi}$ represent the same quantum state; only the relative difference between the orientations of axes is important, as we will see later.

This type of process of finding natural decomposition of the whole Hilbert space into distinguishable `subsystems' from the given dynamics is called quantum mereology \cite{carroll2021quantum}. The decomposition studied in Ref. \cite{carroll2021quantum} was the bipartition into ``system'' and ``environment'' and the criterion was the emergence of quasiclassical behavior. Moreover, there are approaches to explain the emergence of spatial structure from the Hilbert space structure and the given dynamics alone \cite{pollack2019towards, riedel2012rise,cao2017space,van2010building}. The goal of this work is in a similar vein, but the interest of this work is more focused on the emergence of temporal structure in microscopic dynamics, and for doing so we identify the decomposition of the history Hilbert space into future and past systems, which allows the unitary time evolution description of each interaction tensor.

 We will call a tensor (or an operator that can be interpreted as a tensor) $Y\in\mcal{K}$ a \textit{dynamics tensor} if there exists some tensor product decomposition $\mcal{K}=\mcal{K}_1\otimes\mcal{K}_2$ with $\mcal{K}_1\approx\mcal{K}_2$ such that $Y$ is unitary when understood as an operator from $\mcal{K}_1$ to $\mcal{K}_2$. In other words, $Y$ is a dynamics tensor if it is possible to interpret it as a time evolution with respect to some spacetime structure. The following result shows that no special type of tensor network is necessary to represent a dynamics on a pre-spacetime as long as each tensor is properly normalized.

\begin{theorem} \label{thm:dynten}
    Every operator $X\in\mf{B}(\mcal{K})$ is proportional to a dynamics tensor. Especially, every $X$ with $\|X\|_2=|\mcal{K}|^{1/2}$ is a dynamics tensor. 
\end{theorem}
See Appendix for omitted proofs including that of Theorem \ref{thm:dynten}. Treating unitarity as a guideline for assigning temporal order to the pre-spacetime $\mcal{S}$ may help explaining the emergence of temporal structure of spacetime from its tensor network structure, but still there remain some ambiguities. Especially, if $X$ is a \textit{perfect tensor} considered in Ref. \cite{pastawski2015holographic}, then any bipartition of $s_0s_1s_2s_3$ yields a unitary operator, hence it is compatible with any temporal direction direction across the diagram in (\ref{eqn:tennet_1}). Thus, unitarity alone may not yield the unique direction of time.

As we discussed about the unital generalized transpositions and their probable correspondence with temporal axis transformation, now we consider the restriction of the definition of dynamical tensor where the transformation is restricted to unital generalized transpositions. We say that an operator $X\in\mf{B}(\mcal{K})$ is a \textit{proper dynamics tensor} if there exists a unital generalized transposition $T[W]$ with some $W\in\mf{U}(\mcal{K}^{\otimes 2})$ such that $X^{T[W]}$ is unitary.

\begin{theorem} \label{thm:unitdyn}
    Every operator $X\in\mf{B}(\mcal{K})$ is proportional to a proper dynamics tensor if $|\mcal{K}|$ is even. Especially, every $X$ with $\|X\|_2=|\mcal{K}|^{1/2}$ is a proper dynamics tensor. 
\end{theorem}

Theorem \ref{thm:unitdyn} implies that when each subsystem is assumed to be even-dimensional, arbitrary tensor network with properly normalized tensors can be understood as a `net of events' that each constituting tensor can bee seen as a unitary evolution operator after a `rotation of time axis' represented by a unital generalized transposition.

This result lessens the weight of the assumption to justify the approach `spacetime as a tensor network'. According to this viewpoint, there may not be a single universal axis of time in the universe, but each subsystem could experience time as it hops around the given tensor network of dynamics tensors to whatever direction that yields the unitary evolution interpretation for the adjacent dynamics tensor. As there is no unique time axis, each subsystem may `clash' while hopping around the tensor network in different directions. However, this does not necessarily mean that the model is ill-defined or that there is a contradiction, since by satisfying a set of conditions, the interaction between quantum systems with multiple relative configuration of axes of time can be consistent. We will discuss about those conditions in Sec. \ref{subsec:compstp}.

We remark that we do not claim that we found a canonical way to explain the emergence of the unique arrow of time from any tensor network structure of (pre-)spacetime. But we highlight the fact that there could be multiple quantum systems with multiple compatible directions of time in temporally neutral approaches to quantum gravity, and that the generalized transposition provides a mathematical tool to deal with the symmetry transformation of that structure.

This approach bears some similarity with time-symmetric operational formulation of quantum theory such as Oreshkov and Cerf \cite{oreshkov2015operational,oreshkov2016operational} where quantum theory is analyzed without presupposing background spatio-temporal structure. The main difference is that in this work we accept the existence of well-defined local direction of time by interpreting each event as a unitary evolution.

\subsection{Generalized perfect tensors} 
If we were to pursue the approach where a tensor network of events could model the structure of spacetime, then it is natural to expect that the constituting tensors are covariant under a certain class of generalized transposes as we expect physical laws at each point space are `isotropic' in some sense if there is no \textit{ad hoc} temporal axis determined beforehand.

In this Section, we consider continuous generalizations of the class of tensors with particular symmetry known as \textit{perfect tensors}. A perfect tensor $X$ in $\bigotimes_{i=1}^n \mcal{H}_i$ is a tensor which is unitary for any bipartition $A,B$ of $\{1,2,\cdots,n\}$ into input and output nodes with $|A|=|B|$. This definition can be re-expressed in terms of generalized transpositions. We say that $X$ is a perfect tensor, when $X$ is understood as a matrix for some fixed choice of input and output nodes mentioned above, if $X^{T[P]}$ with arbitrary permutation $P$ of $n$ systems is unitary.

Hence, when understood as a tensor describing dynamics in spacetime as in Section \ref{sec:spt}, one can say that a perfect tensor has an `isotropic' spatio-temporal structure to some extent in the sense that even after arbitrary permutation of its nodes it stays unitary. Thus, if one intends to explain the emergence of the spacetime structure from the quantum nature of the universe, then one might think that it is desirable to assume that each dynamics tensor is perfect not to a presupposed temporal structure.

However, as we discussed in Section \ref{sec:spt}, even if we assume that the history Hilbert space of the universe is given as $\mcal{H}=\bigotimes_{s\in\mcal{S}}\mcal{H}_s$, one can always express $\mcal{H}$ with another tensor product structure with some $\mcal{Z}$, i.e. $\mcal{H}=\bigotimes_{z\in\mcal{Z}}\mcal{H}_{z}$. In this regard, the definition of perfect tensor has a shortcoming that it only considers permutation of tensor components of an already fixed tensor product structure of the Hilbert space. This type of property is similar to basis-dependent properties of vectors in the sense it depends on the choice of tensor product structure.

If we were to examine the isotropy of a tensor with respect to arbitrary unitary transformation of Hilbert space, then we must define the following object we will call a \textit{totally perfect tensor}. A tensor $X\in\bigotimes_i\mcal{H}_i\otimes\bigotimes_i\mcal{K}_i$ is totally perfect if (when understood as an operator in $\mf{B}(\bigotimes_i\mcal{H}_i,\bigotimes_i\mcal{K}_i)$) $X^{T[W]}$ is unitary for any unitary operator $W\in\mf{U}(\bigotimes_i\mcal{K}_i\otimes\bigotimes_i\mcal{H}_i)$. Are there totally perfect tensors and do they form isotropic building blocks of spacetime? Or is it the case that there are no such things and the symmetry of choosing regions of spacetime is necessarily broken to some extent? We show that it is the latter.

\begin{theorem} \label{thm:1}
There are no totally perfect tensors.
\end{theorem}
\begin{proof}
    Suppose that $M$ is totally perfect. Then, by letting $W=V(M^\dag\otimes \mds{1})$ for an arbitrary unitary operator $V$, we have $M^{T[W]}=\mds{1}^{T[V]}$. Now, by choosing $V$ such that $\mds{1}^{T[V]}$ is non-unitary, we get the desired result. One such example is the generalized controlled-CNOT gate given as $V=\sum_i S^{-i}\otimes\dyad{i}$, where $S=\sum_n \dyad{n\oplus1}{n}$ where $\oplus$ is the modular summation operation. For such $V$, we have $\mds{1}^{T[V]}=\ket{0}\left(\sum_i\bra{i}\right)$, which is rank-1, hence obviously non-unitary.
\end{proof}

One can understand Theorem \ref{thm:1} as the converse result of Theorem \ref{thm:dynten}. Theorem \ref{thm:1} shows that there is no tensor that is unitary with respect to arbitrary tensor decomposition of the ambient Hilbert space. It implies that every dynamics tensor has a set of preferred (or disfavored, to be more accurate) decomposition of the history Hilbert space, albeit it could still have some remaining ambiguity. This observation helps explaining the emergence of definite pre-spacetime structure as Theorem \ref{thm:1} implies that not every superposition of points in the pre-spacetime can be interpreted as a legitimate subsystem participating in physical interactions.

A naturally following question is if there are operators (tensors) that remain unitary under smaller classes of generalized transpositions. This question is natural since one might guess that there are no totally perfect tensors because the set of all generalized transpositions is too large for an operator to remain unitary after the action of them. We first consider the class of unital generalized transpositions, as they have a good property of preserving zero event. We call an operator $M$ a \textit{properly perfect tensor} if $M^{T[W]}$ is unitary whenever the generalized transposition is unital, i.e. $\mds{1}^{T[W]}=\mds{1}$. We will call the operators of the form $\alpha \mds{1}$ with some complex number $\alpha$ scalar operators.

\begin{prop} \label{prop:noproper}
    There are no properly perfect tensors other than scalar operators.
\end{prop}
\begin{proof}
  Assume that $M \in \mf{B}(A)$ is a properly perfect tensor that is not a scalar operator, i.e., $M\neq \alpha \mds{1}$ for any complex number $\alpha$. Note that $M$ is unitary by definition. Let us decompose $M$ into the trace part and the traceless part. In other words, there exists a traceless operator $S$ with $\|S\|_2=|A|^{1/2}$ that allows for the following form of expansion of $M$
  \begin{equation}
      M=\cos\theta\; \mds{1}_A +\sin\theta S,
  \end{equation}
  for some real value $\theta$ such that $\sin\theta \neq 0$. (Further inspection reveals that $S$ should be unitary, too.) As $S$ is traceless, we can see that $\ket{\phi^+}_{AA'}$ and $(S\otimes\mds{1}_{S'})\ket{\phi^+}_{AA'}$ are orthogonal. This means that one can construct a unitary operator $W$ on $AA'$ that maps $\ket{\phi^+}_{AA'}$ to itself and $(S\otimes\mds{1}_{S'})\ket{\phi^+}_{AA'}$ to $(X\otimes\mds{1}_{S'})\ket{\phi^+}_{AA'}$, where $X=\sum_i \dyad{i\oplus1 (\text{mod } |A|)}{i}$ is the generalized Pauli $X$ operator. Note that $X$ is also traceless. For such $W$, $T[W]$ is unital, hence we have
  \begin{equation}
      N:=M^{T[W]}=\cos\theta\; \mds{1}_A +\sin\theta X.
  \end{equation}
  This operator cannot be unitary since $N^\dag N= \mds{1}_A+ \sin\theta\cos\theta(X+X^\dag)$ and the second term does not vanish. It contradicts the assumption that $\mcal{M}$ is a properly perfect tensor, hence the desired result follows.
\end{proof}

Proposition \ref{prop:noproper} implies that non-scalar dynamics tensors not only have disfavored spacetime  structures as a whole, but also have disfavored temporal structures, if we accept the correspondence between transformations of time axis and unital generalized transpositions.

Nevertheless, as we have seen before, there are unitary operators that are also symmetric, i.e. $U=U^T$ in every dimension. (Note that the direct sum of such operators is also symmetric and unitary.) They are also invariant under fractional transpositions, hence they are all unitary after arbitrary fractional transposition. Hence, we will call an operator $M$ that remain unitary after every fractional transposition, i.e. $M^{T(\theta)}$ is unitary for every $\theta$, a \textit{rotationally perfect tensor} and summarize the observation given above as follows. Although we still lack a complete geometric interpretation of the fractional transposition, one can imagine that the rotationally perfect tensors as tensors that have a time evolution interpretation along with any axis in the space-time plane.

\begin{prop}
    There are rotationally perfect tensors in every dimension.
\end{prop}

\subsection{Supertrace and Factorizalbe Maps} \label{subsec:STr}

In this Section, we introduce a mathematical tool related to the generalized transposition for modelling the loss of \textit{dynamical quantum information} processes. Quantum superchannels are transformations that map quantum channels to quantum channels. Its formal similarity with quantum channels enabled many results about quantum channels to be translated over to quantum superchannels, but not every component of static quantum information processes has been translated into the language of the dynamical setting.

One of such component is information loss. In the static setting, the loss of quantum information is modelled with the (partial) trace operation, and the causality of quantum operation is also formulated in terms of the trace operation. However, to the best of our knowledge, there is no analogue of trace operation for quantum channels, although it is naturally possible that one loses all the information about input and output of a quantum channel.

 We propose the \textit{supertrace} as the superchannel counterpart of the trace operation, denoted by $\mf{Tr}$. The supertrace is defined in such a way that the following diagram is commutative: 
\begin{equation}
    \begin{tikzcd}[baseline=\the\dimexpr\fontdimen22\textfont2\relax]
    \mf{L}(X)\ar[d,"J"]\ar[r,"\mf{Tr}"] & \mds{C} \ar[d,"\text{id}_\mds{C}"] \\
    \mf{B}(X\otimes X')\ar[r,"\Tr"]& \mds{C}
  \end{tikzcd}.
\end{equation}
In other words, $\STr=\mds{J}^{-1}[\Tr]=J^{-1}\circ\Tr\circ J$. Here, we slightly abused the notations by identifying isomorphic trivial Hilbert spaces $\mds{C}^*\approx\mds{C}\approx\mf{L}(\mds{C})\approx\mf{B}(\mds{C}\otimes \mds{C})$ and letting $J:\mf{L}(\mds{C})\to\mf{B}(\mds{C}\otimes\mds{C})$ be identified with $\text{id}_\mds{C}$. Equivalently, $\mf{Tr}[\mcal{M}]:=\Tr[J_{XX'}^\mcal{M}]=\Tr[\mcal{M}(\pi_X)]$ for all $\mcal{M}\in\mf{L}(X)$. Similarly to partial trace, $\STr_X$ is a shorthand expression of $\STr_X \otimes \mf{id}_{\bar{X}}$. It is consistent with the definition of map reduction of Ref. \cite{piani2006properties} where it was defined only for semicausal maps.

Note that the supertrace lacks a few tracial properties such as cyclicity when applied naively, i.e., $\STr[\mcal{A}\circ\mcal{B}]\neq\STr[\mcal{B}\circ\mcal{A}]$ in general on $\mf{L}(X)$. However, it generalizes the operational aspect of trace as the discarding action. For example, every quantum channel $\mcal{N}$ is normalized in supertrace, i.e., $\STr[\mcal{N}]=1$. Not only that, if some linear map $\mcal{M}$ is a quantum channel in \textit{some} configuration, i.e., if $\mcal{M}^{T[W]}$ is a quantum channel for some $W$, then it is still normalized, $\STr[\mcal{M}]=1$. We leave a remark that it is unrelated to the supertrace $\text{Str}$ frequently used in the field of supersymmetry \cite{ma2008supertrace} or the supertrace $\hat{\text{Tr}}$ defined as an operator on endomorphism spaces \cite{caban2002destruction}.

We leave a remark that the normalization condition of quantum processes in Oreshkov and Cerf's generalized process theoretic approach to quantum theory without predefined time \cite{oreshkov2015operational,oreshkov2016operational} can be compactly expressed with supertrace. A quantum operation in their formalism is a CP map $\mcal{M}$ that has unit supertrace: $\STr[\mcal{M}]=1$.

The supertrace yields another way of marginalizing multipartite quantum channels. In other words, we can apply the supertrace to a local system of a bipartite channel to get the quantum channel on the other quantum system. It has advantage over the other definition of marginalization of quantum channel that requires the bipartite channel to be no-signalling \cite{hsieh2022quantum} that the supertrace can be applied to any bipartite channel.

Oftentimes, quantum channels are referred to as deterministic quantum processes in the sense it preserves the trace of input state so that the transformation from the input state to the output state is guaranteed. However, a critical review of its implementation is needed to examine if it can be realized truly deterministically. The Stinespring dilation theorem \cite{stinespring1955positive} implies that for every quantum channel $\mcal{N}\in\mf{C}(X)$, there exists an `ancillary system' $Y$ and a unitary operation $\mcal{U}\in\mf{UO}(XY)$ such that, for every $\rho\in\mf{B}(X)$,
\begin{equation}
    \mcal{N}(\rho)=\Tr_Y \mcal{U}(\rho_X\otimes\dyad{0}_Y),
\end{equation}
for some $\ket{0}$ in $Y$. We have to note that, unless $\mcal{N}$ is a unitary operation, $Y$ is not a 1-dimensional system that only admits $\ket{0}$, but a nontrivial quantum system that can be in some state other than $\ket{0}$. Thus, one role of system $Y$ is providing working space so that information of $X$ can move around to produce the wanted outcome. Another role is providing \textit{purity}. System $Y$ is prepared in a pure state initially, so that the entropy of $X$ can be disposed of. However, how is this pure state $\ket{0}$ prepared? One might claim that the initialization map $\mcal{I}\in\mcal{C}(Y)$ given as
\begin{equation}
    \mcal{I}(\sigma)=\Tr[\sigma]\dyad{0},
\end{equation}
can prepare the pure state $\ket{0}$, but any Stinespring dilation of $\mcal{I}$ itself, for example,
\begin{equation}
    \mcal{I}(\sigma)=\Tr_Z F(\sigma_Y\otimes\dyad{0}_Z)F^\dag,
\end{equation}
with the swapping operator $F$ requires yet another pure ancilla state $\ket{0}_Z$, so the problem will be repeated \textit{ad infinitum}. Indeed,
Landauer's principle asserts that initializing a quantum system inevitably produces heat \cite{landauer1961irreversibility}; entropy can only be displaced, not destroyed, under reversible evolution. The generated heat consumes the purity of the heat absorbent and we again confront the problem of initializing the absorbent.

Another potential solution is preparing the pure state by measuring the ancilla system and choosing the wanted measurement outcomes. However, an operation that can be realized only when some probabilistic measurement outcome happens cannot be deterministic. We also interpret that not just pure states, but any non-maximally mixed quantum state indicates partial knowledge on the given quantum system.

Therefore, by the same argument, every quantum channel that can be deterministically implemented must be possible to realize with a maximally mixed ancilla system, i.e.,
\begin{equation} \label{eqn:stine}
    \mcal{N}(\rho)=\Tr_Y \mcal{U}(\rho_X\otimes\pi_Y).
\end{equation}
Quantum maps of this form are known as \textit{(exactly) factorizable maps}. From the Stinespring dilation of (\ref{eqn:stine}), we can see that any factorizable map $\mcal{M}$ has the following simple expression in terms of supertrace,
\begin{equation} \label{eqn:facpur}
    \mcal{M}=\STr_Y \mcal{U},
\end{equation}
with some unitary operation $\mcal{U}\in\mf{UO}(XY)$. This expression is surprisingly similar to purification of quantum states, i.e. a mixed state $\rho_A$ can be always purified with some environment system $B$ and purification $\ket{\psi}_{AB}$ such that
\begin{equation}
    \rho_A=\Tr_B\dyad{\psi}_{AB}.
\end{equation}
Therefore, we will call $\mcal{U}$ in (\ref{eqn:facpur}) the \textit{purification} of factorizable map $\mcal{M}$. See Appendix \ref{app:factcs} for discussion of general factorizable maps and their relation to general $C^*$-algebras.

By the operational meaning of factorizable maps given here, we can appreciate the physical significance of the mathematical result that not every unital map is factorizable \cite{shor2010structure}. That is, unital maps that are not factorizable require nondeterministic preparation of ancilla systems.

However, the formal similarity of purifications of factorizable maps and quantum states has limitations. For instance, the Schrödinger–HJW theorem \cite{schrodinger1935discussion,hughston1993complete} does not hold for purification of factorizable maps when we try to generalize it straightforwardly.

\begin{prop}
    For some factorizable map $\mcal{N}\in\mf{C}(A)$ and two purifications $\mcal{U}$ and $\mcal{V}$ of $\mcal{N}$ on $AB$, there exists no superunitary operation $\Upsilon \in \mf{SL}(B)$ such that
    \begin{equation}
        \mcal{U}=(\mf{id}_A\otimes\Upsilon_B)(\mcal{V}).
    \end{equation}
\end{prop}
\begin{proof}
    Consider two ways of implementing the completely depolarizing map
\begin{equation}
    \mcal{C}(\rho)=\pi\Tr[\rho].
\end{equation}

 The first method is simply swapping the maximally mixed state with the input state and the second is catalytically depolarizing the input state using a randomness source. The unitary operation of the latter is catalytic; its partial transpose is still unitary and this property does not change after local superunitary. However, the former is not catalytic; the partial transpose of the swapping gate is no longer unitary. Therefore, they cannot be superunitarily similar.
\end{proof} 

Nevertheless, we have the following result immediate from the Schr\"{o}dinger-HJW theorem.
\begin{prop} \label{prop:HJWST}
    For any factorizable map $\mcal{N}\in\mf{C}(A)$ with two purifications $\mcal{U}=\Ad_U$ and $\mcal{V}=\Ad_V$ of $\mcal{N}$ on $AB$, there exists a unitary operator $W \in \mf{U}(B^{\otimes2})$ so that they are related by a generalized transposition $T[W]$, i.e.,
    \begin{equation}
        \mcal{U}=(\mf{id}_A\otimes\mf{T}_B[W])(\mcal{V}).
    \end{equation}
\end{prop}

One possible interpretation of Proposition \ref{prop:HJWST} is that losing dynamical quantum information is not just losing two set of data, namely the input and the output information of a given process, in a fixed temporal structure. Losing dynamical quantum information is symmetric with respect to the transformation of the spatio-temporal structure modelled by generalized transpositions, and it is natural in the sense that there is no \textit{a priori} reason to believe that a quantum system that you have no information at all is governed by the same flow of time with you and the bipartite unitary operator $W$ redirects the temporal progress of the lost system. Indeed, applying a generalized transposition followed by the supertrace is same with applying the supertrace alone, i.e.,
\begin{equation}\label{eqn:sptcaus}
    \mf{Tr}\circ\mf T[W]= \mf{Tr}.
\end{equation}

We remark that (\ref{eqn:sptcaus}) bears a striking similarity with the definition of causality for quantum processes,
\begin{equation}\label{eqn:causality}
    \Tr_Y\circ\;\mcal{E}=\Tr_X.
\end{equation}
Indeed $\mf{T}[W]$ in (\ref{eqn:sptcaus}) simply changing your spacetime-coordinate system for a quantum system about which you have no information at all should not affect all the other processes, hence it expresses a sort of `logical causality'.

\subsection{Compatibility of state preparation} \label{subsec:compstp}
In this Section,  we show that consistency of causal structure is deeply related to flow of information through multipartite interaction, which is greatly important in the study of quantum scrambler, as it was demonstrated in the task of quantum hacking \cite{lie2021hacking}.
As the most evident example, Corollary \ref{coro:maxmix} practically allows no information exchange between two systems compatible with opposite temporal directions. The impossibility of preparing a system compatible with opposite direction of time in a specific state of your choice is evident from the fact that such an action will lead to signaling to the past from the perspective of the inverted system. For example, if a qubit appears to propagate back to the past, then the ability of preparing it in either $\ket{0}$ or $\ket{1}$ is equivalent to the ability of forcing its measurement outcome to be either $\bra{0}$ or $\bra{1}$ on demand in the opposite temporal flow, which leads to retrocausality from that perspective.

In fact, in the quantum setting, bipartite unitary operators $U$ on $AB$ with the unitary partial transpose $U^{T_B}$ are known as \textit{catalytic} unitary operators \cite{lie2020uniform,lie2021correlational,lie2021dynamical} or \textit{T-dual} unitaries \cite{aravinda2021dual}. They allow information exchange between two systems only in the form of randomization given as a unital map \cite{deschamps2016some,benoist2017bipartite}, hence no information leaks to a system initially prepared in the maximally mixed state, as it cannot be randomized more. This is the very reason why these unitary operators can be used for catalysis of quantum randomness \cite{lie2019unconditionally,lie2020randomness,lie2020uniform,lie2021correlational,lie2021dynamical}.

In other words, if you are interacting with a quantum system that is compatible with two opposite temporal flows, then non of your information leaks to it from your perspective. It hints that the more the effect of the given generalized transposition deviates from the usual flow of time, the less information can be leaked through the given interaction.
Henceforth, we can ask the following interesting question: System $A$, a quantum system on your side, is going to interact with another system $B$. You have no knowledge about the interaction between $A$ and $B$ other than it is also compatible with another spatio-temporal structure of $B$ modelled by a generalized transposition $T_B[W]$ on $B$. Does this condition constrain the maximum amount of information that can be leaked from $A$ to $B$?

For this purpose, let us define the (geometric) non-swappability of bipartite quantum channel $\mcal{N}\in\mf{C}(AB)$ with $|A|=|B|$
\begin{equation}
    \Xi(\mcal{N}):=\frac{1}{2}\min_{\mcal{C}_A,\mcal{C}_B}\|\mcal{N}-(\mcal{C}_A\otimes\mcal{C}_B)\circ\Ad_F\|_\diamond,
\end{equation}
where $\mcal{C}_A$ and $\mcal{C}_B$ are local unitary operators. In other words, $\Xi(\mcal{N})$ is the diamond norm distance between $\mcal{N}$ and the set of swapping unitary operations up to local unitary. In this sense, one cay say that $\Xi(\Ad_W)$ is the measure of how close global behavior of $T[W]$ is to the usual transposition.

We also define the \textit{geometric capacity} $C_G(\mcal{N})$ of quantum channel $\mcal{N}$ as
\begin{equation}
    C_G(\mcal{N}):=\frac{1}{2} \min_\tau \|\mcal{N}-\mcal{E}_\tau\|_\diamond,
\end{equation}
where $\mcal{E}_\tau(\rho):=\tau\Tr[\rho]$ from $A$ to $B$ is the initialization map. In other words, $C_G(\mcal{N})$ measures the distance between $\mcal{N}$ and the closest initialization map. State initialization maps completely destroy the information of the input system. Thus, we can say that the farther away a channel is from initialization maps, the more information it preserves.

Hence, when $\cdot \otimes \sigma$ is understood as a quantum channel that attaches an ancilla system in state $\sigma$, we can interpret
\begin{equation}
    \mcal{L}_{B\langle A}(\mcal{M}|\sigma):=C_G(\Tr_A[\mcal{M}(\;\cdot\otimes\sigma)]),
\end{equation}
as the measure of information leakage from $A$ to $B$ for any bipartite channel $\mcal{M}\in\mf{C}(AB)$ when the system $B$ is initially prepared in the state $\sigma$.

\begin{prop} \label{prop:cauloc}
Let $\mcal{U}$ be a bipartite quantum operations on $AB$ compatible with a generalized transposition $T_B[W]$ on $B$, i.e. $\mcal{V}:=\mcal{U}^{T[W^\dag]}$ is also a quantum operation with $\mcal{W}=\Ad_W$. Then, information leakage from $A$ to $B$ through $\mcal{U}$ is limited by the non-swappability of $\mcal{W}$, i.e.,
\begin{equation} \label{eqn:leaklim}
    \max_\sigma \mcal{L}_{B\langle A}(\mcal{U}|\sigma)\leq \Xi(\mcal{W}),
\end{equation}
where the maximization is over quantum states $\sigma$ that are compatible with $T_B[W]$.
\end{prop}
Proof is given in Appendix. One could understand that Proposition \ref{prop:cauloc} provides the robustness of Corollary \ref{coro:maxmix}. For example, if $\mcal{W}$ is the swapping operation corresponding to the ordinary transposition, the right hand side of (\ref{eqn:leaklim}) vanishes, so there could be no information leakage from $A$ to $B$ through $\mcal{U}$. Even when $T[W]$ is slightly different from the ordinary transposition, the information leakage is also very small.

On the other extreme, we examine how highly information leaking interactions restrict the form of compatible partial generalized transposition. We can measure the information destruction by quantum channel $\mcal{N}$ with the minimum sine metric between the Choi matrices of $\mcal{N}$ and an arbitrary unitary operation. Here, the \textit{sine metric} $d_S(\rho,\sigma)$ \cite{rastegin2002relative, rastegin2006sine} between quantum states $\rho$ and $\sigma$ is given as
\begin{equation}
    d_S(\rho,\sigma)=\sqrt{1-F(\rho,\sigma)}.
\end{equation}
Therefore, our (geometric) measure of \textit{information destruction} in $\mcal{N}=\sum_i \Ad_{N_i}\in\mf{C}(A)$ can be expressed as
\begin{equation}
    D_S(\mcal{N}):=\sqrt{1-|A|^{-2}\max_{Y\in\mf{U}(A)} \sum_i\left|\Tr[YN_i]\right|^2}.
\end{equation}
As a special case, $D_S(\mcal{N})$ vanishes if and only if $\mcal{N}$ is a unitary operation that never destroys input information. By using this measure, we can define a geometric measure of \textit{information non-leakage} of bipartite channel $\mcal{M}$ given as 
\begin{equation}
 \mcal{K}_{B\langle A}(\mcal{M}|\sigma):=D_S(\Tr_A[\mcal{M}(\,\cdot\otimes\sigma)]).
\end{equation}
Similarly, we can define the following sine metric-based measure of \textit{non-catalyticity} of bipartite unitary operations,
\begin{equation}
    \mcal{D}_{B\langle A}(\mcal{M}|\sigma):=\min_{\xi\in\mf(B)} d_S(\Tr_A[\mcal{M}(\phi_{AA'}^+\otimes\sigma^T_B)],\pi_{A'}\otimes\xi_B).
\end{equation}
It is a non-catalyticity measure for bipartite unitary operations since, when $\mcal{N}=\Ad_Y$ is a unitary operation, $\mcal{D}_{B\langle A}(\mcal{N}|\sigma)=0$ if and only if $Y$ is a catalytic unitary operator \cite{lie2020uniform, lie2021correlational}.
After preparing these definitions, we can introduce another approximate result on the relation between compatible generalized partial transpositions and information leakage of bipartite quantum channels.
\begin{prop} \label{prop:cauloc2}
Let $\mcal{U}$ be a bipartite unitary operations on $AB$ compatible with a generalized transposition $T_B[W]$ on $B$, i.e. $\mcal{V}:=\mcal{U}^{T[W^\dag]}$ is also a quantum operation with $\mcal{W}=\Ad_W$. Then, non-catalyticity of $\mcal{W}$ is limited by the information non-leakage by $\mcal{U}$, i.e.,
\begin{equation} \label{eqn:leaklim2}
    \mcal{D}_{B'\langle B}(\mcal{W}|\sigma)\leq 2\mcal{K}_{B\langle A}(\mcal{U}|\sigma),
\end{equation}
for all $\sigma$ that are compatible with $T_B[W]$.
\end{prop}
Proof can be found in Appendix.  For example, if $\mcal{U}$ leaks all of the information of input $A$ to output $B$ for a certain initial state $\sigma_B$ of $B$ so that $\mcal{K}_{B\langle A}(\mcal{U}|\sigma)=0$ for some $\sigma$, then any compatible generalized partial transposition $T_B[\mcal{W]}$ must be catalytic, i.e. $\mcal{W}^{T_B}$ is still unitary.

\begin{acknowledgements}
SHL thanks Varun Narasimhachar for insightful discussions. This work was supported by National Research Foundation of Korea grants funded by the Korea government (Grants No. 2019M3E4A1080074, No. 2020R1A2C1008609 and No. 2020K2A9A1A06102946) via the Institute of Applied Physics at Seoul National University and by the Ministry of Science and ICT, Korea, under the ITRC (Information Technology Research Center) support program (IITP-2020-0-01606) supervised by the IITP (Institute of Information \& Communications Technology Planning \& Evaluation) and the quantum computing technology development program of the National Research Foundation of Korea(NRF) funded by the Korean government (Ministry of Science and ICT (MSIT)) (Grants No.2021M3H3A103657312). SHL was also supported by the start-up grant of the Nanyang Assistant Professorship of Nanyang Technological University, Singapore.
\end{acknowledgements}

\appendix
\section{Proof of Theorem 1}
\begin{proof}
    Note that $Y\in\mcal{K}_1\otimes\mcal{K}_2$ being a dynamics tensor is equivalent to the existence of some $W\in\mf{U}(\mcal{K})$ such that $Y^{T[W]}=\mds{1}_{\mcal{K}_1}$ when $Y$ is understood as an operator in $\mf{B}(\mcal{K}_1,\mcal{K}_2)$.    For any two unit vectors in the same Hilbert space, there exists a unitary operator that transforms one to another. Therefore, when $X$ is interpreted as an operator in $\mf(\mcal{K})$, there exists a unitary operator $W\in\mf{U}(\mcal{K}\otimes\mcal{K})$ that transforms $\|X\|_2^{-1}\sum_i X\ket{i}\otimes\ket{i}$ to $ \ket{\phi^+}$. For such $W$, we have $X^{T[W]}=|\mcal{K}|^{-1/2}\|X\|_2\mds{1}_\mcal{K}$.
\end{proof}
\section{Proof of Theorem \ref{thm:unitdyn}}
\begin{proof}
  Suppose that $|\mcal{K}|$ is even. Arbitrary operator $Y\in\mf{B}(\mcal{K})$ has an expansion of the form of $Y=a\mds{1}_\mcal{K}+J$ with a traceless operator $J$. Without loss of generality, we assume that $a$ is real. Note that $\mds{1}_\mcal{K}$ and $J$ are orthogonal to each other. Hence, there exists a unitary operator $W$ on $\mcal{K}^{\otimes 2}$ that preserves $\ket{\phi_{\mcal{K}^{\otimes2}}^+}$ but transforms $(J\otimes\mds{1}_\mcal{K})\ket{\phi_{\mcal{K}^{\otimes2}}^+}$ to $(J'\otimes \mds{1}_\mcal{K})\ket{\phi_{\mcal{K}^{\otimes2}}^+}$ where $J'$ is an arbitrary traceless unitary operator such that $J'^\dag=-J'$, which always exists when $|\mcal{K}|$ is even. An example is direct sum of $i\sigma_Y$ where $\sigma_Y$ is the $2\times 2$ Pauli Y operator. Then, $Y^{T[W]}$ is proportional to a unitary operator as $Y^{T[W]\dag}Y^{T[W]}=|a|^2\mds{1}_\mcal{K} + a(J'+J'^{\dag}) + J'^\dag J= (|a|^2+1)\mds{1}_\mcal{K}$.
\end{proof}
\section{Proof of Theorem \ref{prop:comm}}
\begin{proof}
  For a state preparation superchannel given as $\mf{P}^\sigma(\mcal{N}):=\mcal{N}(\sigma)$ to be compatible with a generalized transposition $T[W]$ of its input channel, equivalently for $\mf{P}^\sigma\circ \mf{T}[W^\dag]$ to be a superchannel, it must be possible to decompose it into
  
  \begin{equation} \label{eqn:comm1}
      \mf{P}^\sigma\circ \mf{T}[W^\dag](\mcal{L})= \Tr_{A'}[\Ad_Q\circ (\mcal{L}_A \otimes \mds{1}_{A'})(\tau_{AA'})],
  \end{equation}
  for any $\mcal{L}\in\mf{L}(A)$, where $Q \in \mf{U}(AA')$ and $\tau_{AA'}$ is a pure quantum state on $AA'$ \cite{chiribella2008transforming}. By applying $\Tr_A$ on the both hands of (\ref{eqn:comm1}) and taking the adjoint (as a map on $\mcal{L}(AA')$) and taking the matrix transposition (as a matrix in $\mcal{L}(AA')$), we get that from the arbitrariness of $\mcal{L}$
  \begin{equation} \label{eqn:comm2}
        W(\mds{1}_{A}\otimes\sigma_{A'}^T)W^\dag=(\mds{1}_{A}\otimes\tau_{A'}^T).
  \end{equation}
  It follows that $\mds{1}_A\otimes\tau_{A'}$ and $\mds{1}_A\otimes \sigma_{A'}$, thus in turn $\tau_A$ and $\sigma_A$ have the same spectrum, therefore they are unitarily similar.
    
    Conversely, if (\ref{eqn:comm2}) holds for some $\tau_A$, then we can set $Q=W$ and $\tau_{AA'}=(\id_{A'}\otimes\Ad_{\sqrt{\tau_A}})(\Gamma_{AA'})$ to express $\mf{P}^\sigma\circ \mf{T}[W^\dag]$ as a superchannel form as in (\ref{eqn:comm1}).
  
\end{proof}
\section{Proof of Proposition \ref{prop:cauloc}}

\begin{proof}
We can observe that, by using the definition of generalized transposition (\ref{eqn:PTe}), $\Tr_A\circ\;\mcal{U}\circ\mcal{A}_\sigma(\rho)$ can be expressed as for any $\rho$,
\begin{equation} \label{eqn:eqex}
        \Tr_{AB'}[(\mds{1}_{AB}\otimes\sigma_{B'}^T)\mcal{W}_{BB'}\circ \mcal{V}_{AB}(\rho_A\otimes\Gamma_{BB'})].
    \end{equation}
However, due to the compatibility of preparing $\sigma$ with $T[W]$, by Proposition \ref{prop:comm} (Note that $W$ and $W^\dag$ are switched in this proof), for the preparation superchannel inputting $\sigma$ to $\mcal{U}$ to be compatible with the transformation $\mcal{U}\mapsto\mf{T}[W^\dag](\mcal{U})$, there must exist a quantum state $\bar{\sigma}$ that is unitarily similar to $\sigma^T$ and satisfies
\begin{equation}
    (\mds{1}_B\otimes\sigma_{B'}^T)W=W(\mds{1}_B\otimes\bar{\sigma}_{B'}).
\end{equation}
Hence, for a purification $\psi_{BB'}=(\id_B\otimes\Ad_{\bar{\sigma}})$ of $\bar{\sigma}_{B'}$, we have that (\ref{eqn:eqex}) equals
    \begin{equation}
        \Tr_{AB'}\circ\mcal{W}_{BB'}\circ \mcal{V}_{AB}(\rho_A\otimes\psi_{BB'}).
    \end{equation}
 Now, we let $\mcal{M}\in\mf{C}(B',B)$ be a channel that achieves
    \begin{equation}
        F_{B\langle B'}(\mcal{W})=\frac{1}{2}\| \Tr_{B'}\circ\;\mcal{W}-\Tr_B\otimes\mcal{M}\|_\diamond.
    \end{equation}
By the submultiplicative property of the diamond norm, as $\|\Tr_A\circ\mcal{V}_{AB}\circ\mcal{A}_\psi\|_\diamond=1$, we have
    \begin{equation}
        \frac{1}{2}\|\Tr_A\circ\;\mcal{U}\circ\mcal{A}_\sigma- \mcal{E}_{\mcal{M}(\bar{\sigma}_{B'})}\|_\diamond \leq F_{B\langle B'}(\mcal{W}).
    \end{equation}
    Here, we used the fact that
    \begin{equation}
        (\Tr_{AB}\otimes\mcal{M})\mcal{V}_{AB}(\rho_A\otimes\psi_{BB'})=\mcal{M}(\bar{\sigma}_{B'}
        )\Tr\rho,
    \end{equation}
    and the definition of $\mcal{E}_\tau$. After the minimization over $\tau$, (\ref{eqn:leaklim}) follows as the choice of $\sigma$ was arbitrary.
\end{proof}

\subsection{Proof of Proposition \ref{prop:cauloc2}}
\begin{proof}
For simplicity, we first assume that $\sigma_B=\pi_B.$ There exist a unitary operator $Y$ on $B$ and, by Uhlmann's theorem, a pure quantum state $\eta_{AB'}$ such that $d_S(\eta_{AB'}\otimes J^{\Ad_Y}_{A'B},J^{\mcal{U}}_{ABA'B'})=\mcal
K_{B\langle A}(\mcal{U}|\pi_B)$. By the monotonicity of fidelity under partial trace, after tracing out $AB$, we have $d_S(\pi_{A'}\otimes \eta_{B'}, \pi_{A'}\otimes\pi_{B'})=d_S(\eta_{B'},\pi_{B'})\leq \mcal{K}_{B\langle A}(\mcal{U}|\pi_B)$. Again, by Uhlmann's theorem, there exists a unitary operator $Z$ such that $d_S(\eta_{AB'},J^{\Ad_Z}_{AB'})=d_S(\eta_{AB'}\otimes J^{\Ad_Y},J^{\Ad_Z}_{AB'}\otimes J^{\Ad_Y})\leq \mcal{K}_{B\langle A}(\mcal{U}|\pi_B)$. Therefore, by the triangle inequality \cite{rastegin2006sine}, we get that $d_S(J^{\Ad_Z}\otimes J^{\Ad_Y},J^{\mcal{U}})$ is upper bounded by
\begin{equation}
    d_S(\eta_{AB'}\otimes J^{\Ad_Y},J^{\mcal{U}}) + d_S(\eta_{AB'}\otimes J^{\Ad_Y},J^{\Ad_Z}\otimes J^{\Ad_Y}), 
\end{equation}
where the subscripts for the Choi matrices are omitted. Both terms are upper bounded by $\mcal{K}_{B\langle A}(\mcal{U}|\pi_B)$. Again, by the monotonicity of fidelity, by applying $\Tr_{AB}\circ\mcal{W}_{BB'}$ to both $J^{\Ad_Z}\otimes J^{\Ad_Y}$ and $J^{\mcal{U}}$, we get
\begin{equation}
    d_S(\Tr_B[\mcal{W}_{BB'}(\phi^+_{A'B}\otimes\pi_{B'})],\pi_{A'B'})\leq 2\mcal{K}_{B\langle A}(\mcal{U}|\pi_B).
\end{equation}
Now, observe that the left hand side is $\mcal{D}_{B\langle B'}(\mcal{W}|\pi_{B'})$.

For general $\sigma_B$, the proof is more or less similar except for that $J^{\mcal{U}}_{ABA'B'}$ is replaced with $\zeta_{ABA'B'}:=(
\id_{A'B'}\otimes\mcal{U}_{AB})(\phi^+_{AA'}\otimes\sigma_{BB'})$, where $\sigma_{BB'}:=\Ad_{\sqrt{\sigma_B}}(\phi^+_{BB'})$ is a purification of the given $\sigma_B$. Then there exist $\eta_{AB'}$ and some unitary $Y$ such that $d_S(\eta_{AB'}\otimes J^{\Ad_{Y}}_{A'B},\zeta_{ABA'B'})=\mcal{K}_{B\langle A}(\mcal{U}|\sigma_B)$ and $d_S(\eta_{B'},\sigma_{B'})\leq\mcal{K}_{B\langle A}(\mcal{U}|\sigma_B)$. Note that $\sigma_{B'}=\sigma^T_B$. As we did for $\sigma_B=\pi_B$ case, we apply $\mcal{W}$ on $BB'$ and trace out $AB$ to both $\zeta_{ABA'B'}$ and $J^{\Ad_Y}_{A'B}\otimes\eta_{AB'}$. By using the compatibility condition (\ref{eqn:prepcom}) and that $\mcal{U}^{T[W]}=\mcal{V}$, we get that the former turns into $\pi_{A'}\otimes\tau^T_{B'}$ for some $\tau_B$ and the latter is mapped to $\Tr_B[\mcal{W}(J^{\Ad_Y}_{A'B}\otimes \eta_{B'})]$. Note that the sine metric betwee $\Tr_B[\mcal{W}(J^{\Ad_Y}_{A'B}\otimes \eta_{B'})]$ and $\Tr_B[\mcal{W}(J^{\Ad_Y}_{A'B}\otimes \sigma_{B'})]$ is upper bounded by $d_S(\eta_{B'},\sigma_{B'})\leq\mcal{K}_{B\langle A}(\mcal{U}|\sigma_B)$. Since $d_S(\pi_{A'}\otimes\tau_{B'},\Tr_B[\mcal{W}](J^{\Ad_Y}_{A'B}\otimes \sigma_{B'}))$ bounds $\mcal{D}_S(\mcal{W}|\sigma)$ from above, by using the triangle inequality of the sine metric, we get the wanted result.
\end{proof}

\section{Factorizable maps with general $C^*$-algebra} \label{app:factcs}
In contrast to the fact that postselecting a certain measurement outcome of a system whose state can be changed when affected but has not yet been examined before is not deterministic, we claim that generating a state with randomness which cannot be altered afterwards is deterministically implementable. This is because, in that case, we assume that there is no room for the change of the ancilla system, so no measurement is required to identify its state, and we are not selecting a certain subset of states but using the whole probabilistically mixed state.

Hence, we can also deterministically implement the aforementioned exactly factorizable maps conditioned on the classical register. This more general set of quantum maps are known as \textit{factorizable} maps and can be expressed as follows with some probability distribution $\{p_i\}$,

Since $|Y|^{-1}\sum_i p_i \Tr_{Y_i}$ is a tracial state of $C^*$-algebra $\bigoplus_i \mf{B}(Y_i)$ for every probability distribution $\{p_i\}$, by just naming it $\Tr_Y:=\sum_i p_i \Tr_{Y_i}$ so that $\STr_Y:=\sum_i p_i \STr_{Y_i}$, we again recover the purification expression for an arbitrary factorizable map (\ref{eqn:facpur}).

\begin{equation}
    \mcal{N}(\rho)=\sum_i p_i \Tr_{Y_i} \mcal{U}_i(\rho_X\otimes \pi_{Y_i}).
\end{equation}
Here, $Y$ is decomposed into superselection sectors $Y=\bigoplus_i Y_i$ with the orthogonal projector $\Pi_i$ onto each subspace $Y_i$, and $\mcal{U}_i\in\mf{UO}(X,Y_i)$. Note that this is a simpler expression for finite dimensional $Y$, but the concept of factorizable maps can be also defined on infinite dimensional systems. (See \cite{haagerup2011factorization,musat2020non} for more information.) We focus on finite dimensional cases in this work for simplicity.

\bibliographystyle{unsrtnat}
\bibliography{main}

\end{document}

%% file: sample.tikzstyles
\tikzstyle{red dot}=[fill=red, draw=black, shape=circle]
\tikzstyle{green dot}=[fill={rgb,255: red,0; green,128; blue,128}, draw=black, shape=circle]
\tikzstyle{vertical box}=[fill=white, draw=black, shape=rectangle, minimum width=0.75cm, minimum height=1cm]
\tikzstyle{horizontal box}=[fill=white, draw=black, shape=rectangle, minimum width=1cm, minimum height=0.75cm]
\tikzstyle{th-v}=[fill=white, draw=black, shape=rectangle, minimum width=0.5cm, minimum height=1cm, tikzit category=thin box]
\tikzstyle{th-h}=[fill=white, draw=black, shape=rectangle, tikzit category=thin box, minimum width=1cm, minimum height=0.5cm]
\tikzstyle{SQ}=[fill=white, draw=black, shape=rectangle, minimum width=0.5cm, minimum height=0.5cm]
\tikzstyle{white}=[fill=white, draw=black, shape=circle]

\tikzstyle{thick white}=[-, line width=2pt, draw=white]

%% file: trans.tikz
\begin{tikzpicture}
	\begin{pgfonlayer}{nodelayer}
		\node [style=SQ] (0) at (-2, 0) {$M^T$};
		\node [style=none] (1) at (-3.25, 0) {};
		\node [style=none] (2) at (-0.75, 0) {};
		\node [style=none] (3) at (-0.25, 0) {$=$};
		\node [style=SQ] (5) at (1.25, 0) {$M$};
		\node [style=none] (6) at (1.5, -1) {};
		\node [style=none] (7) at (1, 1) {};
		\node [style=none] (8) at (2, 1) {};
		\node [style=none] (9) at (0.5, -1) {};
		\node [style=none] (10) at (1.5, 0) {};
		\node [style=none] (11) at (1, 0) {};
	\end{pgfonlayer}
	\begin{pgfonlayer}{edgelayer}
		\draw (1.center) to (0);
		\draw (0) to (2.center);
		\draw (7.center) to (8.center);
		\draw (9.center) to (6.center);
		\draw [bend right=90, looseness=2.00] (7.center) to (11.center);
		\draw [bend left=90, looseness=2.00] (10.center) to (6.center);
	\end{pgfonlayer}
\end{tikzpicture}

%% file: TWtrans.tikz
\begin{tikzpicture}
	\begin{pgfonlayer}{nodelayer}
		\node [style=none] (0) at (1.25, 0) {};
		\node [style=none] (3) at (0.5, 0) {};
		\node [style=none] (4) at (3, 0) {};
		\node [style=none] (7) at (2.25, 0) {};
		\node [style=none] (12) at (3.75, 0) {:=};
		\node [style=none] (13) at (5.25, 1) {};
		\node [style=none] (16) at (5, 0.75) {};
		\node [style=none] (23) at (5, -0.75) {};
		\node [style=none] (24) at (6.5, -0.75) {};
		\node [style=none] (25) at (4.5, -1.25) {};
		\node [style=none] (26) at (7, -1.25) {};
		\node [style=none] (30) at (5, 0.25) {};
		\node [style=SQ] (38) at (1.75, 0) {$M^T$};
		\node [style=SQ] (39) at (5.75, 1) {$M$};
		\node [style=none] (40) at (6.25, 1) {};
		\node [style=none] (41) at (6.5, 0.75) {};
		\node [style=none] (42) at (6.5, 0.25) {};
		\node [style=none] (43) at (5.75, -0.25) {};
		\node [style=none] (44) at (5, -0.75) {};
		\node [style=none] (45) at (6.5, 0.25) {};
		\node [style=none] (46) at (5.75, -0.25) {};
	\end{pgfonlayer}
	\begin{pgfonlayer}{edgelayer}
		\draw (3.center) to (0.center);
		\draw (7.center) to (4.center);
		\draw [bend left=45, looseness=1.50] (16.center) to (13.center);
		\draw [bend left=45, looseness=1.25] (23.center) to (25.center);
		\draw [bend right=45] (24.center) to (26.center);
		\draw [in=90, out=-90] (16.center) to (30.center);
		\draw [bend right=45, looseness=1.50] (41.center) to (40.center);
		\draw (41.center) to (42.center);
		\draw [in=180, out=-90, looseness=0.75] (30.center) to (43.center);
		\draw [in=90, out=0, looseness=0.75] (43.center) to (24.center);
		\draw [style=thick white, in=0, out=-90, looseness=0.75] (42.center) to (43.center);
		\draw [style=thick white, in=90, out=180, looseness=0.75] (43.center) to (23.center);
		\draw [in=0, out=-90, looseness=0.75] (45.center) to (46.center);
		\draw [in=90, out=180, looseness=0.75] (46.center) to (44.center);
	\end{pgfonlayer}
\end{tikzpicture}

%% file: TWident.tikz
\begin{tikzpicture}
	\begin{pgfonlayer}{nodelayer}
		\node [style=none] (0) at (1.75, 0) {};
		\node [style=none] (3) at (1, 0) {};
		\node [style=none] (4) at (3, 0) {};
		\node [style=none] (7) at (2.25, 0) {};
		\node [style=none] (12) at (3.75, 0) {:=};
		\node [style=none] (13) at (5, 0.5) {};
		\node [style=none] (16) at (4.75, 0.25) {};
		\node [style=none] (23) at (4.75, -0.5) {};
		\node [style=none] (24) at (6.25, -0.5) {};
		\node [style=none] (25) at (4.25, -1) {};
		\node [style=none] (26) at (6.75, -1) {};
		\node [style=none] (30) at (4.75, -0.5) {};
		\node [style=SQ] (38) at (2, 0) {$M$};
		\node [style=SQ] (39) at (5.5, 0.5) {$M$};
		\node [style=none] (40) at (6, 0.5) {};
		\node [style=none] (41) at (6.25, 0.25) {};
		\node [style=none] (42) at (6.25, -0.5) {};
		\node [style=none] (44) at (4.75, -0.5) {};
	\end{pgfonlayer}
	\begin{pgfonlayer}{edgelayer}
		\draw (3.center) to (0.center);
		\draw (7.center) to (4.center);
		\draw [bend left=45, looseness=1.50] (16.center) to (13.center);
		\draw [bend left=45, looseness=1.25] (23.center) to (25.center);
		\draw [bend right=45] (24.center) to (26.center);
		\draw [in=90, out=-90] (16.center) to (30.center);
		\draw [bend right=45, looseness=1.50] (41.center) to (40.center);
		\draw (41.center) to (42.center);
	\end{pgfonlayer}
\end{tikzpicture}

%% file: TW0.tikz
\begin{tikzpicture}
	\begin{pgfonlayer}{nodelayer}
		\node [style=none] (0) at (0, 0) {};
		\node [style=none] (3) at (-0.5, 0) {};
		\node [style=none] (4) at (3, 0) {};
		\node [style=none] (7) at (2.5, 0) {};
		\node [style=none] (12) at (3.75, 0) {:=};
		\node [style=none] (13) at (5.25, 1) {};
		\node [style=none] (16) at (5, 0.75) {};
		\node [style=th-h] (22) at (5.75, -0.5) {$W$};
		\node [style=none] (23) at (5, -1) {};
		\node [style=none] (24) at (6.5, -1) {};
		\node [style=none] (25) at (4.5, -1.5) {};
		\node [style=none] (26) at (7, -1.5) {};
		\node [style=none] (30) at (5, 0) {};
		\node [style=none] (34) at (1.25, 0.75) {{\tiny$\leftarrow$}};
		\node [style=none] (36) at (4.5, -0.5) {{\tiny$\downarrow$}};
		\node [style=none] (37) at (5.75, 1.75) {{\tiny$\leftarrow$}};
		\node [style=SQ] (38) at (1.25, 0) {$M^{T[W]}$};
		\node [style=SQ] (39) at (5.75, 1) {$M$};
		\node [style=none] (40) at (6.25, 1) {};
		\node [style=none] (41) at (6.5, 0.75) {};
		\node [style=none] (42) at (6.5, 0) {};
	\end{pgfonlayer}
	\begin{pgfonlayer}{edgelayer}
		\draw (3.center) to (0.center);
		\draw (7.center) to (4.center);
		\draw [bend left=45, looseness=1.50] (16.center) to (13.center);
		\draw [bend left=45, looseness=1.25] (23.center) to (25.center);
		\draw [bend right=45] (24.center) to (26.center);
		\draw [in=90, out=-90] (16.center) to (30.center);
		\draw [bend right=45, looseness=1.50] (41.center) to (40.center);
		\draw (41.center) to (42.center);
	\end{pgfonlayer}
\end{tikzpicture}

%% file: TWpart.tikz
\begin{tikzpicture}
	\begin{pgfonlayer}{nodelayer}
		\node [style=none] (0) at (-0.25, 0) {};
		\node [style=none] (3) at (-1, 0) {};
		\node [style=none, label={above:$B$}] (4) at (3, 0) {};
		\node [style=none] (7) at (2.25, 0) {};
		\node [style=none] (12) at (3.75, 0) {:=};
		\node [style=none] (13) at (5.25, 1) {};
		\node [style=none] (16) at (5, 0.75) {};
		\node [style=th-h] (22) at (5.75, -0.5) {$W$};
		\node [style=none] (23) at (5, -1) {};
		\node [style=none] (24) at (6.5, -1) {};
		\node [style=none] (25) at (4.5, -1.5) {};
		\node [style=none, label={above:$B$}] (26) at (7, -1.5) {};
		\node [style=none] (30) at (5, 0) {};
		\node [style=none] (34) at (1, 1.75) {{\tiny$\leftarrow$}};
		\node [style=none] (36) at (4.5, -0.5) {{\tiny$\downarrow$}};
		\node [style=none] (37) at (5.75, 2.75) {{\tiny$\leftarrow$}};
		\node [style=th-v] (38) at (1, 0.5) {$M^{T_B[W]}$};
		\node [style=th-v] (39) at (5.75, 1.5) {$M$};
		\node [style=none] (40) at (6.25, 1) {};
		\node [style=none] (41) at (6.5, 0.75) {};
		\node [style=none] (42) at (6.5, 0) {};
		\node [style=none] (43) at (-0.25, 1) {};
		\node [style=none] (44) at (-1, 1) {};
		\node [style=none, label={above:$A$}] (45) at (3, 1) {};
		\node [style=none] (46) at (2.25, 1) {};
		\node [style=none] (47) at (5.25, 2) {};
		\node [style=none] (48) at (4.75, 2) {};
		\node [style=none, label={above:$A$}] (49) at (6.75, 2) {};
		\node [style=none] (50) at (6.25, 2) {};
	\end{pgfonlayer}
	\begin{pgfonlayer}{edgelayer}
		\draw (3.center) to (0.center);
		\draw (7.center) to (4.center);
		\draw [bend left=45, looseness=1.50] (16.center) to (13.center);
		\draw [bend left=45, looseness=1.25] (23.center) to (25.center);
		\draw [bend right=45] (24.center) to (26.center);
		\draw [in=90, out=-90] (16.center) to (30.center);
		\draw [bend right=45, looseness=1.50] (41.center) to (40.center);
		\draw (41.center) to (42.center);
		\draw (44.center) to (43.center);
		\draw (46.center) to (45.center);
		\draw (48.center) to (47.center);
		\draw (50.center) to (49.center);
	\end{pgfonlayer}
\end{tikzpicture}

%% file: tennet_1.tikz
\begin{tikzpicture}
	\begin{pgfonlayer}{nodelayer}
		\node [style=none] (0) at (0, 0.5) {};
		\node [style=none] (3) at (-1.25, 0) {};
		\node [style=none] (43) at (0, 0.5) {};
		\node [style=none] (44) at (-1.25, 1) {};
		\node [style=none] (46) at (0, 0.5) {};
		\node [style=none] (47) at (1.75, 1) {$s_0$};
		\node [style=none] (48) at (1.75, 0) {$s_1$};
		\node [style=none] (49) at (-2, 1) {$s_2$};
		\node [style=none] (50) at (-2, 0) {$s_3$};
		\node [style=none] (51) at (0, 0.5) {};
		\node [style=none] (52) at (0, 0.5) {};
		\node [style=none] (53) at (1.25, 0) {};
		\node [style=none] (54) at (0, 0.5) {};
		\node [style=none] (55) at (1.25, 1) {};
		\node [style=none] (56) at (0, 0.5) {};
		\node [style=white] (57) at (0, 0.5) {$X$};
	\end{pgfonlayer}
	\begin{pgfonlayer}{edgelayer}
		\draw [in=-90, out=0, looseness=0.50] (3.center) to (0.center);
		\draw [in=90, out=0, looseness=0.50] (44.center) to (43.center);
		\draw [in=-90, out=180, looseness=0.50] (53.center) to (52.center);
		\draw [in=90, out=180, looseness=0.50] (55.center) to (54.center);
	\end{pgfonlayer}
\end{tikzpicture}

%% file: main.bbl
\begin{thebibliography}{67}
\providecommand{\natexlab}[1]{#1}
\providecommand{\url}[1]{\texttt{#1}}
\expandafter\ifx\csname urlstyle\endcsname\relax
  \providecommand{\doi}[1]{doi: #1}\else
  \providecommand{\doi}{doi: \begingroup \urlstyle{rm}\Url}\fi

\bibitem[Chiribella and Liu(2022)]{chiribella2020quantum}
Giulio Chiribella and Zixuan Liu.
\newblock Quantum operations with indefinite time direction.
\newblock \emph{Communications Physics}, 5\penalty0 (1):\penalty0 190, 2022.
\newblock \doi{https://doi.org/10.1038/s42005-022-00967-3}.

\bibitem[Chiribella et~al.(2013)Chiribella, D’Ariano, Perinotti, and
  Valiron]{chiribella2013quantum}
Giulio Chiribella, Giacomo~Mauro D’Ariano, Paolo Perinotti, and Benoit
  Valiron.
\newblock Quantum computations without definite causal structure.
\newblock \emph{Physical Review A}, 88\penalty0 (2):\penalty0 022318, 2013.
\newblock \doi{https://doi.org/10.1103/PhysRevA.88.022318}.

\bibitem[Page and Wootters(1983)]{page1983evolution}
Don~N Page and William~K Wootters.
\newblock Evolution without evolution: Dynamics described by stationary
  observables.
\newblock \emph{Physical Review D}, 27\penalty0 (12):\penalty0 2885, 1983.
\newblock \doi{https://doi.org/10.1103/PhysRevD.27.2885}.

\bibitem[Albrecht and Iglesias(2008)]{albrecht2008clock}
Andreas Albrecht and Alberto Iglesias.
\newblock Clock ambiguity and the emergence of physical laws.
\newblock \emph{Physical Review D}, 77\penalty0 (6):\penalty0 063506, 2008.
\newblock \doi{https://doi.org/10.1103/PhysRevD.77.063506}.

\bibitem[Albrecht and Iglesias(2012)]{albrecht2012clock}
Andreas Albrecht and Alberto Iglesias.
\newblock The clock ambiguity: Implications and new developments.
\newblock In \emph{The Arrows of Time}, pages 53--68. Springer, 2012.
\newblock \doi{https://doi.org/10.1007/978-3-642-23259-6_4}.

\bibitem[Marletto and Vedral(2017)]{marletto2017evolution}
Chiara Marletto and Vlatko Vedral.
\newblock Evolution without evolution and without ambiguities.
\newblock \emph{Physical Review D}, 95\penalty0 (4):\penalty0 043510, 2017.
\newblock \doi{https://doi.org/10.1103/PhysRevD.95.043510}.

\bibitem[Choi(1975)]{choi1975completely}
Man-Duen Choi.
\newblock Completely positive linear maps on complex matrices.
\newblock \emph{Linear algebra and its applications}, 10\penalty0 (3):\penalty0
  285--290, 1975.
\newblock \doi{https://doi.org/10.1016/0024-3795(75)90075-0}.

\bibitem[Jamio{\l}kowski(1972)]{jamiolkowski1972linear}
Andrzej Jamio{\l}kowski.
\newblock Linear transformations which preserve trace and positive
  semidefiniteness of operators.
\newblock \emph{Reports on Mathematical Physics}, 3\penalty0 (4):\penalty0
  275--278, 1972.
\newblock \doi{https://doi.org/10.1016/0034-4877(72)90011-0}.

\bibitem[Chiribella et~al.(2008)Chiribella, D'Ariano, and
  Perinotti]{chiribella2008transforming}
Giulio Chiribella, Giacomo~Mauro D'Ariano, and Paolo Perinotti.
\newblock Transforming quantum operations: Quantum supermaps.
\newblock \emph{EPL (Europhysics Letters)}, 83\penalty0 (3):\penalty0 30004,
  2008.
\newblock \doi{https://doi.org/10.1209/0295-5075/83/30004}.

\bibitem[Gour(2019)]{gour2019comparison}
Gilad Gour.
\newblock Comparison of quantum channels by superchannels.
\newblock \emph{IEEE Transactions on Information Theory}, 65\penalty0
  (9):\penalty0 5880--5904, 2019.
\newblock \doi{https://doi.org/10.1109/TIT.2019.2907989}.

\bibitem[Chiribella et~al.(2009)Chiribella, D’Ariano, and
  Perinotti]{chiribella2009theoretical}
Giulio Chiribella, Giacomo~Mauro D’Ariano, and Paolo Perinotti.
\newblock Theoretical framework for quantum networks.
\newblock \emph{Physical Review A}, 80\penalty0 (2):\penalty0 022339, 2009.
\newblock \doi{https://doi.org/10.1103/PhysRevA.80.022339}.

\bibitem[Burniston et~al.(2020)Burniston, Grabowecky, Scandolo, Chiribella, and
  Gour]{burniston2020necessary}
John Burniston, Michael Grabowecky, Carlo~Maria Scandolo, Giulio Chiribella,
  and Gilad Gour.
\newblock Necessary and sufficient conditions on measurements of quantum
  channels.
\newblock \emph{Proceedings of the Royal Society A}, 476\penalty0
  (2236):\penalty0 20190832, 2020.

\bibitem[Bisio and Perinotti(2019)]{bisio2019theoretical}
Alessandro Bisio and Paolo Perinotti.
\newblock Theoretical framework for higher-order quantum theory.
\newblock \emph{Proceedings of the Royal Society A}, 475\penalty0
  (2225):\penalty0 20180706, 2019.
\newblock \doi{https://doi.org/10.1098/rspa.2018.0706}.

\bibitem[Gour and Scandolo(2020)]{gour2020dynamical}
Gilad Gour and Carlo~Maria Scandolo.
\newblock Dynamical resources.
\newblock \emph{arXiv preprint arXiv:2101.01552}, 2020.

\bibitem[Peres(1980)]{peres1980measurement}
Asher Peres.
\newblock Measurement of time by quantum clocks.
\newblock \emph{American Journal of Physics}, 48\penalty0 (7):\penalty0
  552--557, 1980.
\newblock \doi{https://doi.org/10.1119/1.12061}.

\bibitem[Bu{\v{z}}ek et~al.(1999)Bu{\v{z}}ek, Derka, and
  Massar]{buvzek1999optimal}
Vladimir Bu{\v{z}}ek, Radoslav Derka, and Serge Massar.
\newblock Optimal quantum clocks.
\newblock \emph{Physical review letters}, 82\penalty0 (10):\penalty0 2207,
  1999.
\newblock \doi{https://doi.org/10.1103/PhysRevLett.82.2207}.

\bibitem[Aharonov et~al.(1990)Aharonov, Anandan, Popescu, and
  Vaidman]{aharonov1990superpositions}
Yakir Aharonov, Jeeva Anandan, Sandu Popescu, and Lev Vaidman.
\newblock Superpositions of time evolutions of a quantum system and a quantum
  time-translation machine.
\newblock \emph{Physical review letters}, 64\penalty0 (25):\penalty0 2965,
  1990.
\newblock \doi{https://doi.org/10.1103/PhysRevLett.64.2965}.

\bibitem[Ball(2017)]{ball2017world}
Philip Ball.
\newblock A world without cause and effect.
\newblock \emph{Nature}, 546\penalty0 (29):\penalty0 590--92, 2017.
\newblock \doi{https://doi.org/10.1038/546590a}.

\bibitem[Horodecki et~al.(1996)Horodecki, Horodecki, and
  Horodecki]{horodecki1996necessary}
Michal Horodecki, Pawel Horodecki, and Ryszard Horodecki.
\newblock On the necessary and sufficient conditions for separability of mixed
  quantum states.
\newblock \emph{Phys. Lett. A}, 223\penalty0 (1), 1996.
\newblock \doi{https://doi.org/10.1016/S0375-9601(96)00706-2}.

\bibitem[Ye et~al.(2004)Ye, Sun, Zhang, and Guo]{ye2004entanglement}
Ming-Yong Ye, Dong Sun, Yong-Sheng Zhang, and Guang-Can Guo.
\newblock Entanglement-changing power of two-qubit unitary operations.
\newblock \emph{Physical Review A}, 70\penalty0 (2):\penalty0 022326, 2004.
\newblock \doi{https://doi.org/10.1103/PhysRevA.70.022326}.

\bibitem[Kraus and Cirac(2001)]{kraus2001optimal}
Barbara Kraus and Juan~I Cirac.
\newblock Optimal creation of entanglement using a two-qubit gate.
\newblock \emph{Physical Review A}, 63\penalty0 (6):\penalty0 062309, 2001.
\newblock \doi{https://doi.org/10.1103/PhysRevA.63.062309}.

\bibitem[Pucha{\l}a et~al.(2015)Pucha{\l}a, Rudnicki, Chabuda, Paraniak, and
  {\.Z}yczkowski]{puchala2015certainty}
Zbigniew Pucha{\l}a, {\L}ukasz Rudnicki, Krzysztof Chabuda, Miko{\l}aj
  Paraniak, and Karol {\.Z}yczkowski.
\newblock Certainty relations, mutual entanglement, and nondisplaceable
  manifolds.
\newblock \emph{Physical Review A}, 92\penalty0 (3):\penalty0 032109, 2015.
\newblock \doi{https://doi.org/10.1103/PhysRevA.92.032109}.

\bibitem[Brahmachari et~al.(2022)Brahmachari, Rajmohan, Rather, and
  Lakshminarayan]{brahmachari2022dual}
Shrigyan Brahmachari, Rohan~Narayan Rajmohan, Suhail~Ahmad Rather, and Arul
  Lakshminarayan.
\newblock Dual unitaries as maximizers of the distance to local product gates.
\newblock \emph{arXiv preprint arXiv:2210.13307}, 2022.

\bibitem[Shor et~al.(2022{\natexlab{a}})Shor, Benninger, and
  Khrennikov]{shor2022towards}
Oded Shor, Felix Benninger, and Andrei Khrennikov.
\newblock Towards unification of general relativity and quantum theory:
  Dendrogram representation of the event-universe.
\newblock \emph{Entropy}, 24\penalty0 (2):\penalty0 181, 2022{\natexlab{a}}.
\newblock \doi{https://doi.org/10.3390/e24020181}.

\bibitem[Shor et~al.(2022{\natexlab{b}})Shor, Benninger, and
  Khrennikov]{shor2022dendrographic}
Oded Shor, Felix Benninger, and Andrei Khrennikov.
\newblock Dendrographic hologram theory: Predictability of relational dynamics
  of the event universe and the emergence of time arrow.
\newblock \emph{Symmetry}, 14\penalty0 (6):\penalty0 1089, 2022{\natexlab{b}}.
\newblock \doi{https://doi.org/10.3390/sym14061089}.

\bibitem[Cotler(2020)]{cotler2020toward}
Jordan Cotler.
\newblock \emph{Toward the Emergence of Time in Quantum Gravity}.
\newblock Stanford University, 2020.

\bibitem[Cotler and Wilczek(2016)]{cotler2016entangled}
Jordan Cotler and Frank Wilczek.
\newblock Entangled histories.
\newblock \emph{Physica Scripta}, 2016\penalty0 (T168):\penalty0 014004, 2016.
\newblock \doi{https://doi.org/10.1088/0031-8949/2016/T168/014004}.

\bibitem[Castellani(2021{\natexlab{a}})]{castellani2021entropy}
Leonardo Castellani.
\newblock Entropy of temporal entanglement.
\newblock \emph{arXiv preprint arXiv:2104.05722}, 2021{\natexlab{a}}.

\bibitem[Castellani(2021{\natexlab{b}})]{castellani2021history}
Leonardo Castellani.
\newblock History entanglement entropy.
\newblock \emph{Physica Scripta}, 96\penalty0 (5):\penalty0 055217,
  2021{\natexlab{b}}.
\newblock \doi{https://doi.org/10.1088/1402-4896/abe6c0}.

\bibitem[Dias(2021)]{dias2021quantum}
Eduardo~O Dias.
\newblock Quantum formalism for events and how time can emerge from its
  foundations.
\newblock \emph{Physical Review A}, 103\penalty0 (1):\penalty0 012219, 2021.
\newblock \doi{https://doi.org/10.1103/PhysRevA.103.012219}.

\bibitem[Isham et~al.(2003)]{isham2003new}
CJ~Isham et~al.
\newblock A new approach to quantising space-time: I. quantising on a general
  category.
\newblock \emph{Advances in Theoretical and Mathematical Physics}, 7\penalty0
  (2):\penalty0 331--367, 2003.
\newblock \doi{https://doi.org/10.4310/ATMP.2003.v7.n2.a5}.

\bibitem[Cotler et~al.(2018)Cotler, Jian, Qi, and
  Wilczek]{cotler2018superdensity}
Jordan Cotler, Chao-Ming Jian, Xiao-Liang Qi, and Frank Wilczek.
\newblock Superdensity operators for spacetime quantum mechanics.
\newblock \emph{Journal of High Energy Physics}, 2018\penalty0 (9):\penalty0
  1--57, 2018.
\newblock \doi{https://doi.org/10.1007/JHEP09(2018)093}.

\bibitem[Farhi and Gutmann(1992)]{farhi1992functional}
Edward Farhi and Sam Gutmann.
\newblock The functional integral constructed directly from the hamiltonian.
\newblock \emph{Annals of Physics}, 213\penalty0 (1):\penalty0 182--203, 1992.
\newblock \doi{https://doi.org/10.1016/0003-4916(92)90288-W}.

\bibitem[Carroll and Singh(2019)]{carroll2019mad}
Sean~M Carroll and Ashmeet Singh.
\newblock Mad-dog everettianism: Quantum mechanics at its most minimal.
\newblock In \emph{What is Fundamental?}, pages 95--104. Springer, 2019.
\newblock \doi{https://doi.org/10.1007/978-3-030-11301-8_10}.

\bibitem[Carroll and Singh(2021)]{carroll2021quantum}
Sean~M Carroll and Ashmeet Singh.
\newblock Quantum mereology: Factorizing hilbert space into subsystems with
  quasiclassical dynamics.
\newblock \emph{Physical Review A}, 103\penalty0 (2):\penalty0 022213, 2021.
\newblock \doi{https://doi.org/10.1103/PhysRevA.103.022213}.

\bibitem[Cao et~al.(2017)Cao, Carroll, and Michalakis]{cao2017space}
ChunJun Cao, Sean~M Carroll, and Spyridon Michalakis.
\newblock Space from hilbert space: recovering geometry from bulk entanglement.
\newblock \emph{Physical Review D}, 95\penalty0 (2):\penalty0 024031, 2017.
\newblock \doi{https://doi.org/10.1103/PhysRevD.95.024031}.

\bibitem[Hosur et~al.(2016)Hosur, Qi, Roberts, and Yoshida]{hosur2016chaos}
Pavan Hosur, Xiao-Liang Qi, Daniel~A Roberts, and Beni Yoshida.
\newblock Chaos in quantum channels.
\newblock \emph{Journal of High Energy Physics}, 2016\penalty0 (2):\penalty0
  1--49, 2016.

\bibitem[Pastawski et~al.(2015)Pastawski, Yoshida, Harlow, and
  Preskill]{pastawski2015holographic}
Fernando Pastawski, Beni Yoshida, Daniel Harlow, and John Preskill.
\newblock Holographic quantum error-correcting codes: Toy models for the
  bulk/boundary correspondence.
\newblock \emph{Journal of High Energy Physics}, 2015\penalty0 (6):\penalty0
  1--55, 2015.
\newblock \doi{https://doi.org/10.1109/MPUL.2015.2397294}.

\bibitem[May(2017)]{may2017tensor}
Alex May.
\newblock Tensor networks for dynamic spacetimes.
\newblock \emph{Journal of High Energy Physics}, 2017\penalty0 (6):\penalty0
  1--24, 2017.
\newblock \doi{https://doi.org/10.1007/JHEP06(2017)118}.

\bibitem[Allen et~al.(2017)Allen, Barrett, Horsman, Lee, and
  Spekkens]{allen2017quantum}
John-Mark~A Allen, Jonathan Barrett, Dominic~C Horsman, Ciar{\'a}n~M Lee, and
  Robert~W Spekkens.
\newblock Quantum common causes and quantum causal models.
\newblock \emph{Physical Review X}, 7\penalty0 (3):\penalty0 031021, 2017.
\newblock \doi{https://doi.org/10.1103/PhysRevX.7.031021}.

\bibitem[Pollack and Singh(2019)]{pollack2019towards}
Jason Pollack and Ashmeet Singh.
\newblock Towards space from hilbert space: finding lattice structure in
  finite-dimensional quantum systems.
\newblock \emph{Quantum Studies: Mathematics and Foundations}, 6\penalty0
  (2):\penalty0 181--200, 2019.
\newblock \doi{https://doi.org/10.1007/JHEP02(2016)004}.

\bibitem[Riedel et~al.(2012)Riedel, Zurek, and Zwolak]{riedel2012rise}
C~Jess Riedel, Wojciech~H Zurek, and Michael Zwolak.
\newblock The rise and fall of redundancy in decoherence and quantum darwinism.
\newblock \emph{New Journal of physics}, 14\penalty0 (8):\penalty0 083010,
  2012.
\newblock \doi{https://doi.org/10.1088/1367-2630/14/8/083010}.

\bibitem[Van~Raamsdonk(2010)]{van2010building}
Mark Van~Raamsdonk.
\newblock Building up spacetime with quantum entanglement.
\newblock \emph{General Relativity and Gravitation}, 42\penalty0 (10):\penalty0
  2323--2329, 2010.
\newblock \doi{https://doi.org/10.1007/s10714-010-1034-0}.

\bibitem[Oreshkov and Cerf(2015)]{oreshkov2015operational}
Ognyan Oreshkov and Nicolas~J Cerf.
\newblock Operational formulation of time reversal in quantum theory.
\newblock \emph{Nature Physics}, 11\penalty0 (10):\penalty0 853--858, 2015.
\newblock \doi{https://doi.org/10.1038/nphys3414}.

\bibitem[Oreshkov and Cerf(2016)]{oreshkov2016operational}
Ognyan Oreshkov and Nicolas~J Cerf.
\newblock Operational quantum theory without predefined time.
\newblock \emph{New Journal of Physics}, 18\penalty0 (7):\penalty0 073037,
  2016.
\newblock \doi{https://doi.org/10.1088/1367-2630/18/7/073037}.

\bibitem[Piani et~al.(2006)Piani, Horodecki, Horodecki, and
  Horodecki]{piani2006properties}
Marco Piani, Michal Horodecki, Pawel Horodecki, and Ryszard Horodecki.
\newblock Properties of quantum nonsignaling boxes.
\newblock \emph{Physical Review A}, 74\penalty0 (1):\penalty0 012305, 2006.
\newblock \doi{https://doi.org/10.1103/PhysRevA.74.012305}.

\bibitem[Ma et~al.(2008)Ma, He, and Qin]{ma2008supertrace}
Wen-Xiu Ma, Jing-Song He, and Zhen-Yun Qin.
\newblock A supertrace identity and its applications to superintegrable
  systems.
\newblock \emph{Journal of Mathematical Physics}, 49\penalty0 (3):\penalty0
  033511, 2008.
\newblock \doi{https://doi.org/10.1063/1.2897036}.

\bibitem[Caban et~al.(2002)Caban, Rembieli{\'n}ski, Smoli{\'n}ski, and
  Walczak]{caban2002destruction}
P~Caban, J~Rembieli{\'n}ski, KA~Smoli{\'n}ski, and Z~Walczak.
\newblock Destruction of states in quantum mechanics.
\newblock \emph{Journal of Physics A: Mathematical and General}, 35\penalty0
  (14):\penalty0 3265, 2002.
\newblock \doi{https://doi.org/10.1088/0305-4470/35/14/308}.

\bibitem[Hsieh et~al.(2022)Hsieh, Lostaglio, and Ac{\'\i}n]{hsieh2022quantum}
Chung-Yun Hsieh, Matteo Lostaglio, and Antonio Ac{\'\i}n.
\newblock Quantum channel marginal problem.
\newblock \emph{Physical Review Research}, 4\penalty0 (1):\penalty0 013249,
  2022.
\newblock \doi{https://doi.org/10.1103/PhysRevResearch.4.013249}.

\bibitem[Stinespring(1955)]{stinespring1955positive}
W~Forrest Stinespring.
\newblock Positive functions on c*-algebras.
\newblock \emph{Proceedings of the American Mathematical Society}, 6\penalty0
  (2):\penalty0 211--216, 1955.
\newblock \doi{https://doi.org/10.1090/S0002-9939-1955-0069403-4}.

\bibitem[Landauer(1961)]{landauer1961irreversibility}
Rolf Landauer.
\newblock Irreversibility and heat generation in the computing process.
\newblock \emph{IBM journal of research and development}, 5\penalty0
  (3):\penalty0 183--191, 1961.
\newblock \doi{https://doi.org/10.1147/rd.53.0183}.

\bibitem[Shor(2010)]{shor2010structure}
Peter Shor.
\newblock Structure of unital maps and the asymptotic quantum birkhoff
  conjecture, 2010.

\bibitem[Schr{\"o}dinger(1935)]{schrodinger1935discussion}
Erwin Schr{\"o}dinger.
\newblock Discussion of probability relations between separated systems.
\newblock In \emph{Mathematical Proceedings of the Cambridge Philosophical
  Society}, volume~31, pages 555--563. Cambridge University Press, 1935.
\newblock \doi{https://doi.org/10.1017/S0305004100013554}.

\bibitem[Hughston et~al.(1993)Hughston, Jozsa, and
  Wootters]{hughston1993complete}
Lane~P Hughston, Richard Jozsa, and William~K Wootters.
\newblock A complete classification of quantum ensembles having a given density
  matrix.
\newblock \emph{Physics Letters A}, 183\penalty0 (1):\penalty0 14--18, 1993.
\newblock \doi{https://doi.org/10.1016/0375-9601(93)90880-9}.

\bibitem[Lie et~al.(2021)Lie, Teo, and Jeong]{lie2021hacking}
Seok~Hyung Lie, Yong~Siah Teo, and Hyunseok Jeong.
\newblock Hacking quantum networks: Extraction and installation of quantum
  data.
\newblock \emph{arXiv preprint arXiv:2105.13823}, 2021.

\bibitem[Lie and Jeong(2021{\natexlab{a}})]{lie2020uniform}
Seok~Hyung Lie and Hyunseok Jeong.
\newblock Randomness for quantum channels: Genericity of catalysis and quantum
  advantage of uniformness.
\newblock \emph{Physical Review Research}, 3\penalty0 (1):\penalty0 013218,
  2021{\natexlab{a}}.
\newblock \doi{https://doi.org/10.1103/PhysRevResearch.3.013218}.

\bibitem[Lie and Jeong(2021{\natexlab{b}})]{lie2021correlational}
Seok~Hyung Lie and Hyunseok Jeong.
\newblock Catalytic quantum randomness as a correlational resource.
\newblock \emph{Physical Review Research}, 3\penalty0 (4):\penalty0 043089,
  2021{\natexlab{b}}.
\newblock \doi{https://doi.org/10.1103/PhysRevResearch.3.043089}.

\bibitem[Lie and Jeong(2023)]{lie2021dynamical}
Seok~Hyung Lie and Hyunseok Jeong.
\newblock Delocalized and dynamical catalytic randomness and information flow.
\newblock \emph{Physical Review A}, 107\penalty0 (4):\penalty0 042430, 2023.
\newblock \doi{https://doi.org/10.1103/PhysRevA.107.042430}.

\bibitem[Aravinda et~al.(2021)Aravinda, Rather, and
  Lakshminarayan]{aravinda2021dual}
S~Aravinda, Suhail~Ahmad Rather, and Arul Lakshminarayan.
\newblock From dual-unitary to quantum bernoulli circuits: Role of the
  entangling power in constructing a quantum ergodic hierarchy.
\newblock \emph{arXiv preprint arXiv:2101.04580}, 2021.

\bibitem[Deschamps et~al.(2016)Deschamps, Nechita, and
  Pellegrini]{deschamps2016some}
Julien Deschamps, Ion Nechita, and Cl{\'e}ment Pellegrini.
\newblock On some classes of bipartite unitary operators.
\newblock \emph{Journal of Physics A: Mathematical and Theoretical},
  49\penalty0 (33):\penalty0 335301, 2016.
\newblock \doi{https://doi.org/10.1088/1751-8113/49/33/335301}.

\bibitem[Benoist and Nechita(2017)]{benoist2017bipartite}
Tristan Benoist and Ion Nechita.
\newblock On bipartite unitary matrices generating subalgebra-preserving
  quantum operations.
\newblock \emph{Linear Algebra and its Applications}, 521:\penalty0 70--103,
  2017.
\newblock \doi{https://doi.org/10.1016/j.laa.2017.01.020}.

\bibitem[Lie et~al.(2019)Lie, Kwon, Kim, and Jeong]{lie2019unconditionally}
Seok~Hyung Lie, Hyukjoon Kwon, MS~Kim, and Hyunseok Jeong.
\newblock Unconditionally secure qubit commitment scheme using quantum maskers.
\newblock \emph{arXiv preprint arXiv:1903.12304}, 2019.

\bibitem[Lie and Jeong(2020)]{lie2020randomness}
Seok~Hyung Lie and Hyunseok Jeong.
\newblock Randomness cost of masking quantum information and the information
  conservation law.
\newblock \emph{Physical Review A}, 101\penalty0 (5):\penalty0 052322, 2020.
\newblock \doi{https://doi.org/10.1103/PhysRevA.101.052322}.

\bibitem[Rastegin(2002)]{rastegin2002relative}
AE~Rastegin.
\newblock Relative error of state-dependent cloning.
\newblock \emph{Physical Review A}, 66\penalty0 (4):\penalty0 042304, 2002.
\newblock \doi{https://doi.org/10.1103/PhysRevA.66.042304}.

\bibitem[Rastegin(2006)]{rastegin2006sine}
Alexey~E Rastegin.
\newblock Sine distance for quantum states.
\newblock \emph{arXiv preprint quant-ph/0602112}, 2006.
\newblock \doi{https://doi.org/10.48550/arXiv.quant-ph/0602112}.

\bibitem[Haagerup and Musat(2011)]{haagerup2011factorization}
Uffe Haagerup and Magdalena Musat.
\newblock Factorization and dilation problems for completely positive maps on
  von neumann algebras.
\newblock \emph{Communications in Mathematical Physics}, 303\penalty0
  (2):\penalty0 555--594, 2011.
\newblock \doi{https://doi.org/10.1007/s00220-011-1216-y}.

\bibitem[Musat and R{\o}rdam(2020)]{musat2020non}
Magdalena Musat and Mikael R{\o}rdam.
\newblock Non-closure of quantum correlation matrices and factorizable channels
  that require infinite dimensional ancilla (with an appendix by narutaka
  ozawa).
\newblock \emph{Communications in Mathematical Physics}, 375\penalty0
  (3):\penalty0 1761--1776, 2020.
\newblock \doi{https://doi.org/10.1007/s00220-019-03449-w}.

\end{thebibliography}
